\documentclass[prd,aps,nofootinbib,floatfix,11pt]{revtex4}
\usepackage{amsmath,graphicx,epsfig,amssymb,dsfont,mathtools,}
\usepackage[usenames]{color}
\usepackage{ulem} 
\usepackage{bigstrut}
\usepackage{slashed}
\usepackage{multirow}

\allowdisplaybreaks

\begin{document}
\title{QCD Sum Rules Analysis of Weak Decays of Doubly-Heavy Baryons}
\author{Yu-Ji Shi$^{1}$~\footnote{Corresponding author, Email:shiyuji@sjtu.edu.cn}, Wei Wang$^{1}$~\footnote{Corresponding author, Email:wei.wang@sjtu.edu.cn},
and Zhen-Xing Zhao$^{1}$~\footnote{Corresponding author, Email:star\_0027@sjtu.edu.cn}}
\affiliation{$^{1}$ INPAC, SKLPPC,
School of Physics and Astronomy, Shanghai Jiao Tong University, Shanghai 200240, China}
\begin{abstract}
We calculate the weak decay form factors of doubly-heavy baryons  using  three-point QCD sum rules.  The Cutkosky rules are used to derive the double dispersion relations.  We include perturbative contributions and condensation  contributions up to dimension five, and point out that   the perturbative contributions and   condensates with lowest dimensions dominate. An estimate of part of gluon-gluon condensates  show that it plays a less important role. With these form factors at hand, we present a phenomenological study of semileptonic decays. The future experimental facilities can test these predictions, and deepen our understanding of the dynamics in decays of doubly-heavy baryons. 
\end{abstract}
\maketitle

\section{Introduction}

Although quark model has achieved many brilliant successes in hadron spectroscopy, not all predicted  particles, even in ground-state,  in the quark model have been experimentally established so far. These states include  doubly-heavy baryons and triply-heavy baryons. 
In 2017, the LHCb collaboration has reported the  first observation of doubly-charmed baryon $\Xi_{cc}^{++}$ with the mass~\cite{Aaij:2017ueg}
\begin{equation}
m_{\Xi_{cc}^{++}}=(3621.40\pm0.72\pm0.27\pm0.14)\ {\rm MeV}\label{eq:LHCb_measurement}
\end{equation}
in the $\Lambda_{c}^{+}K^{-}\pi^{+}\pi^{+}$ final state.
Soon afterwards  new results on  $\Xi_{cc}^{++}$ were released by LHCb, including  the first measurement of its lifetime~\cite{Aaij:2018wzf} and the observation of a new decay mode $\Xi_{cc}^{++}\to\Xi_{c}^{+}\pi^{+}$~\cite{Aaij:2018gfl}. On   experimental side, more investigations on $\Xi_{cc}^{++}$ and searches for other doubly-heavy baryons are certainly demanded to achieve a better understanding~\cite{Traill:2017zbs,Cerri:2018ypt}.  
Meanwhile  these observations have triggered many theoretical studies  on various properties of  doubly-heavy baryons~\cite{Wang:2017mqp,Meng:2017udf,Wang:2017azm,
Gutsche:2017hux,Li:2017pxa,Guo:2017vcf,Lu:2017meb,Xiao:2017udy,Sharma:2017txj,Ma:2017nik,Yu:2017zst,Meng:2017dni,Hu:2017dzi,Cui:2017udv,Shi:2017dto,Xiao:2017dly,Yao:2018zze,Yao:2018ifh,
Ozdem:2018uue,Ali:2018ifm,Dias:2018qhp,Li:2018epz,Zhao:2018mrg,Xing:2018bqt,Zhu:2018epc,
Ali:2018xfq,Liu:2018euh,Xing:2018lre,Bediaga:2018lhg,Wang:2018duy,Dhir:2018twm,Berezhnoy:2018bde,Jiang:2018oak,Zhang:2018llc,Li:2018bkh,Meng:2018zbl,Cerri:2018ypt,Gutsche:2018msz}, most of which have been focused on the spectrum, production and decay properties.

In a previous work~\cite{Wang:2017mqp}, we have performed  an analysis of decay form factors of doubly-heavy baryons  in a light-front quark model (LFQM). In this light-front study, the diquark
picture is adopted, where the two spectator quarks are treated  as a
bounded system.  This approximation  can greatly simplify  the calculation and many useful phenomenological results are obtained~\cite{Zhao:2018mrg,Xing:2018lre}.  But meanwhile this diquark approximation introduces uncontrollable systematic uncertainties since the dynamics in the diquark system has been smeared. In this work,  we will remedy this shortcoming and perform an analysis  of transition form factors using QCD sum rules (QCDSR). Some earlier attempts basing on non-relativistic QCD (NRQCD) sum rules can be found in Refs.~\cite{Onishchenko:2000wf,Onishchenko:2000yp,Kiselev:2001fw}. It is necessary to note that since the decay final state contains only one heavy quark, NRQCD should not be applicable unless the strange quark is also treated as a heavy quark. In the literature   the QCDSR framework has also been used to calculate  masses and the pole residues of doubly heavy baryons in a number of references (see for instance~\cite{Zhang:2008rt,Wang:2010hs,Wang:2010vn,Wang:2010it,Hu:2017dzi}). So it is desirable to calculate the decay form factors within the same framework, which is the motif of this work. 

\begin{figure}
\includegraphics[width=0.5\columnwidth]{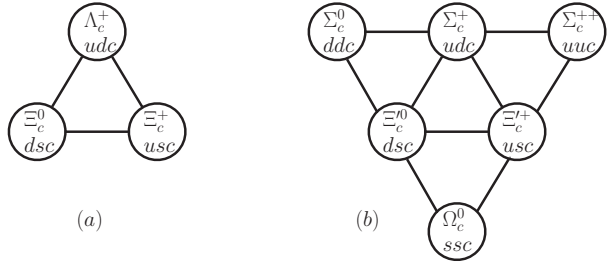} \caption{The anti-triplet (panel a) and sextet (panel b) of charmed baryons. It is similar for bottom baryons. }
\label{fig:one_heavy} 
\end{figure}

In our analysis, the doubly heavy baryons  include  $\Xi_{cc}(ccq)$, $\Omega_{cc}(ccs)$, $\Xi_{bb}(bbq)$, $\Omega_{bb}(bbs)$, and $\Xi_{bc}(bcq)$, $\Omega_{bc}(bcs)$, with $q=u,d$. The  $\Xi_{QQ^{\prime}}$ and $\Omega_{QQ^{\prime}}$ can form a flavor  SU(3) triplet.
It should be noted that the two heavy quarks in $\Xi_{bc}$ and $\Omega_{bc}$ are symmetric in the flavor space. The antisymmetric case that    presumably will decay via strong or electromagnetic interactions are not   considered in this work. Quantum numbers of  doubly heavy baryons can be found in Table~\ref{tab:JPC}. Baryons in the final state contains one heavy bottom/charm quark and two light quarks. They can form an SU(3) anti-triplet $\Lambda_{Q}$,
$\Xi_{Q}$ or an SU(3) sextet $\Sigma_{Q}$, $\Xi_{Q}^{\prime}$ and $\Omega_{Q}$ with $Q=b,c$, as depicted  in  Fig.~\ref{fig:one_heavy}.  
	
\begin{table*}[!htb]
\caption{Quantum numbers and quark content for the lowest-lying doubly heavy baryons. $S_{h}^{\pi}$ denotes the spin/parity of the system of two heavy quarks. The light quark $q$ corresponds to the $u,d$ quark. } \label{tab:JPC}  
\begin{tabular}{cccc|cccccc}
\hline 
{\footnotesize{}{}{}Baryon }  & {\footnotesize{}{}{}Quark content }  & {\footnotesize{}{}{}$S_{h}^{\pi}$ }  & {\footnotesize{}{}{}$J^{P}$ }  & {\footnotesize{}{}{}Baryon }  & {\footnotesize{}{}{}Quark content }  & {\footnotesize{}{}{}$S_{h}^{\pi}$ }  & {\footnotesize{}{}{}$J^{P}$ }  &  & \tabularnewline
			\hline 
			{\footnotesize{}{}{}$\Xi_{cc}$ }  & {\footnotesize{}{}{}$\{cc\}q$ }  & {\footnotesize{}{}{}$1^{+}$ }  & {\footnotesize{}{}{}$1/2^{+}$ }  & {\footnotesize{}{}{}$\Xi_{bb}$ }  & {\footnotesize{}{}{}$\{bb\}q$ }  & {\footnotesize{}{}{}$1^{+}$ }  & {\footnotesize{}{}{}$1/2^{+}$ }  &  & \tabularnewline
			{\footnotesize{}{}{}$\Xi_{cc}^{*}$ }  & {\footnotesize{}{}{}$\{cc\}q$ }  & {\footnotesize{}{}{}$1^{+}$ }  & {\footnotesize{}{}{}$3/2^{+}$ }  & {\footnotesize{}{}{}$\Xi_{bb}^{*}$ }  & {\footnotesize{}{}{}$\{bb\}q$ }  & {\footnotesize{}{}{}$1^{+}$ }  & {\footnotesize{}{}{}$3/2^{+}$ }  &  & \tabularnewline
			\hline 
			{\footnotesize{}{}{}$\Omega_{cc}$ }  & {\footnotesize{}{}{}$\{cc\}s$ }  & {\footnotesize{}{}{}$1^{+}$ }  & {\footnotesize{}{}{}$1/2^{+}$ }  & {\footnotesize{}{}{}$\Omega_{bb}$ }  & {\footnotesize{}{}{}$\{bb\}s$ }  & {\footnotesize{}{}{}$1^{+}$ }  & {\footnotesize{}{}{}$1/2^{+}$ }  &  & \tabularnewline
			{\footnotesize{}{}{}$\Omega_{cc}^{*}$ }  & {\footnotesize{}{}{}$\{cc\}s$ }  & {\footnotesize{}{}{}$1^{+}$ }  & {\footnotesize{}{}{}$3/2^{+}$ }  & {\footnotesize{}{}{}$\Omega_{bb}^{*}$ }  & {\footnotesize{}{}{}$\{bb\}s$ }  & {\footnotesize{}{}{}$1^{+}$ }  & {\footnotesize{}{}{}$3/2^{+}$ }  &  & \tabularnewline
			\hline 
			{\footnotesize{}{}{}$\Xi_{bc}^{\prime}$ }  & {\footnotesize{}{}{}$[bc]q$ }  & {\footnotesize{}{}{}$0^{+}$ }  & {\footnotesize{}{}{}$1/2^{+}$ }  & {\footnotesize{}{}{}$\Omega_{bc}^{\prime}$ }  & {\footnotesize{}{}{}$[bc]s$ }  & {\footnotesize{}{}{}$0^{+}$ }  & {\footnotesize{}{}{}$1/2^{+}$ }  &  & \tabularnewline
			{\footnotesize{}{}{}$\Xi_{bc}$ }  & {\footnotesize{}{}{}$\{bc\}q$ }  & {\footnotesize{}{}{}$1^{+}$ }  & {\footnotesize{}{}{}$1/2^{+}$ }  & {\footnotesize{}{}{}$\Omega_{bc}$ }  & {\footnotesize{}{}{}$\{bc\}s$ }  & {\footnotesize{}{}{}$1^{+}$ }  & {\footnotesize{}{}{}$1/2^{+}$ }  &  & \tabularnewline
			{\footnotesize{}{}{}$\Xi_{bc}^{*}$ }  & {\footnotesize{}{}{}$\{bc\}q$ }  & {\footnotesize{}{}{}$1^{+}$ }  & {\footnotesize{}{}{}$3/2^{+}$ }  & {\footnotesize{}{}{}$\Omega_{bc}^{*}$ }  & {\footnotesize{}{}{}$\{bc\}s$ }  & {\footnotesize{}{}{}$1^{+}$ }  & {\footnotesize{}{}{}$3/2^{+}$ }  &  & \tabularnewline
			\hline 
	\end{tabular}
	\end{table*}

To be more explicit, the transitions of doubly heavy baryons  can be  classified as follows: 
\begin{itemize}
\item The $cc$ sector 
\begin{eqnarray}
\Xi_{cc} & \to&[\Lambda_{c},\Xi_{c},\Sigma_{c},\Xi_{c}^{\prime}], \;\;\;
\Omega_{cc}  \to[\Xi_{c},\Xi_{c}^{\prime}],\nonumber
\end{eqnarray}
\item The $bb$ sector 
		\begin{eqnarray}
		\Xi_{bb} & \to&[\Lambda_{b},\Sigma_{b}],\;\;\;
		\Omega_{bb} \to[\Xi_{b},\Xi_{b}^{\prime}],\nonumber
\end{eqnarray}
\item The $bc$ sector with $c$ quark decay 
		\begin{eqnarray}
		\Xi_{bc} & \to&[\Lambda_{b},\Xi_{b},\Sigma_{b},\Xi_{b}^{\prime}],\;\;\;
		\Omega_{bc} \to[\Xi_{b},\Xi_{b}^{\prime}],\nonumber
		\end{eqnarray}
		\item The $bc$ sector with $b$ quark decay 
		\begin{eqnarray}
		\Xi_{bc} & \to&[\Lambda_{c},\Sigma_{c}],\;\;\;
		\Omega_{bc} \to[\Xi_{c},\Xi_{c}^{\prime}]. \nonumber
		\end{eqnarray}
	\end{itemize}
In the above, both  SU(3) anti-triplet and sextet  final states are taken into account. However, the $b\to c$ transition will not be considered in this work, and is left for future.

The rest of this paper is arranged as follows. In Sec.~\ref{sec:sum_rules},  the transition form factors are calculated in QCDSR, where the perturbative contribution, quark condensates, quark-gluon condensates are calculated and an estimate of part of gluon-gluon condensates is presented. Numerical results for form factors are presented in Sec.~\ref{sec:numerical}, which are subsequently  used to perform  the phenomenological studies  in Sec.~\ref{sec:phenomenological}. A brief summary  of this work and the prospect for the future are given in the last section.
	
\section{Transition Form Factors in QCD sum rules}
\label{sec:sum_rules}

\subsection{Form Factors}

\begin{figure}
\includegraphics[width=0.4\columnwidth]{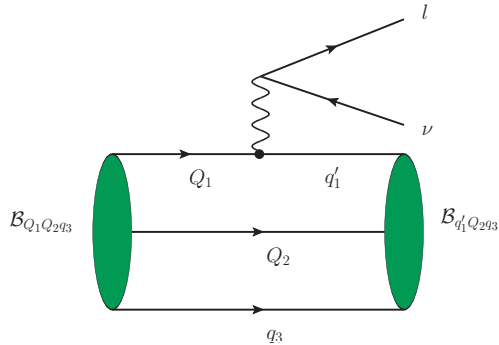} 
\caption{Feynman diagram for semileptonic decays. The leptonic amplitude can be calculated using perturbation theory, while hadronic matrix elements can be parametrized into form factors.   }
\label{fig:semi_lepton} 
\end{figure}

We show the Feynman diagram for semileptonic decays of doubly-heavy baryons in Fig.~\ref{fig:semi_lepton}.  
The leptonic amplitude in this transition  can be calculated using electro-weak perturbation theory, while the hadronic matrix elements can be parametrized into transition form factors:  
\begin{eqnarray}
\langle{\cal B}_{2}(p_{2})|(V-A)_{\mu}|{\cal B}_{1}(p_{1})\rangle  & = & \bar{u}(p_{2},s_{2})\bigg[\gamma_{\mu}f_{1}(q^{2})+i\sigma_{\mu\nu}\frac{q^{\nu}}{M_{1}}f_{2}(q^{2})+\frac{q_{\mu}}{M_{1}}f_{3}(q^{2})\bigg]u(p_{1},s_{1})\nonumber \\
&   &-\bar{u}(p_{2},s_{2})\bigg[\gamma_{\mu}g_{1}(q^{2})+i\sigma_{\mu\nu}\frac{q^{\nu}}{M_{1}}g_{2}(q^{2})+\frac{q_{\mu}}{M_{1}}g_{3}(q^{2})\bigg]\gamma_{5}u(p_{1},s_{1}), 
\label{eq:parameterization}
\end{eqnarray}
where $p_1(s_1)$ is the momentum (spin) of the initial state, and $p_2(s_2)$ is the momentum (spin) of the final baryon.  The momentum transfer is defined as $q^\mu= p_1^\mu -p_2^\mu$, and the vector (axial-vector) $V^\mu(A^\mu)$ is defined as $\bar q_1' \gamma^\mu (\gamma^\mu\gamma^5) Q_1$, with $q_1'$ being a light quark and $Q_1$ as a heavy bottom or charm quark.  $M_1$ is the mass of the initial doubly-heavy baryon. These form factors are also responsible for non-leptonic decay modes if the factorization holds, and thus  must be calculated  in a nonperturbative manner for later use. 

In our calculation, the following simple parametrization will be used first:
\begin{eqnarray}
\langle{\cal B}_{2}(p_{2},s_{2})|(V-A)_{\mu}|{\cal B}_{1}(p_{1},s_{1})\rangle & = & \bar{u}(p_{2},s_{2})[\frac{p_{1\mu}}{M_{1}}F_{1}+\frac{p_{2\mu}}{M_{2}}F_{2}+\gamma_{\mu}F_{3}]u(p_{1},s_{1})\nonumber \\
 & - & \bar{u}(p_{2},s_{2})[\frac{p_{1\mu}}{M_{1}}G_{1}+\frac{p_{2\mu}}{M_{2}}G_{2}+\gamma_{\mu}G_{3}]\gamma_{5}u(p_{1},s_{1}).
 \label{eq:parameterization_2}
\end{eqnarray}
Once the form factors $F_{i}$ and $G_{i}$ in Eq. (\ref{eq:parameterization_2}) are obtained, then they will be transformed into $f_{i}$ and $g_{i}$ in Eq. (\ref{eq:parameterization}), which are used to compared with other works in the literature.

\subsection{QCD Sum Rules}

The starting point in QCDSR is to construct a suitable correlation function, and for the   ${\cal B}_{Q_{1}Q_{2}q_{3}}\to{\cal B}_{q_{1}^{\prime}Q_{2}q_{3}}$ transition, it is chosen as: 
\begin{equation}
	\Pi_{\mu}^{V,A}(p_{1}^{2},p_{2}^{2},q^{2})=i^{2}\int d^{4}xd^{4}ye^{-ip_{1}\cdot x+ip_{2}\cdot y}\langle0|T\{J_{{\cal B}_{q_{1}^{\prime}Q_{2}q_{3}}}(y)(V_{\mu},A_\mu)(0)\bar{J}_{{\cal B}_{Q_{1}Q_{2}q_{3}}}(x)\}|0\rangle.  \label{eq:corrfunction}
\end{equation} 
Here the weak transition $Q_{1}\to q_{1}^{\prime}$ stands for the  $c\to d/s$ or $b\to u$ process. The $Q_{2}=c/b$, $q_{3}=u/d/s$ and $V_\mu(A_{\mu})=\bar{q}_{1}^{\prime}\gamma_{\mu}(\gamma_\mu\gamma_{5})Q_{1}$.
The  $J_{{\cal B}_{q_{1}^{\prime}Q_{2}q_{3}}}$ and $J_{{\cal B}_{Q_{1}Q_{2}q_{3}}}$ are the interpolating currents for singly and doubly heavy baryons respectively.  For $\Xi_{QQ}$ and $\Omega_{QQ}$, they are used as: 
\begin{align}
J_{\Xi_{QQ}} & =\epsilon_{abc}(Q_{a}^{T}C\gamma^{\mu}Q_{b})\gamma_{\mu}\gamma_{5}q_{c},\nonumber \\
J_{\Omega_{QQ}} & =\epsilon_{abc}(Q_{a}^{T}C\gamma^{\mu}Q_{b})\gamma_{\mu}\gamma_{5}s_{c},\label{eq:current_QQ}
\end{align}
where $Q=b,c$ and $q=u,d$. For $\Xi_{bc}$ and $\Omega_{bc}$ the interpolating currents are 
\begin{align}
J_{\Xi_{bc}} & =\frac{1}{\sqrt{2}}\epsilon_{abc}(b_{a}^{T}C\gamma^{\mu}c_{b}+c_{a}^{T}C\gamma^{\mu}b_{b})\gamma_{\mu}\gamma_{5}q_{c},\nonumber \\
J_{\Omega_{bc}} & =\frac{1}{\sqrt{2}}\epsilon_{abc}(b_{a}^{T}C\gamma^{\mu}c_{b}+c_{a}^{T}C\gamma^{\mu}b_{b})\gamma_{\mu}\gamma_{5}s_{c},\label{eq:current_bc}
\end{align}
where $b$ and $c$ fields are chosen symmetric. The interpolating currents for singly heavy baryons can be defined in a similar way. For the SU(3) anti-triplet they are 
\begin{eqnarray}
J_{\Lambda_{Q}} & = & \frac{1}{\sqrt{2}}\epsilon_{abc}(u_{a}^{T}C\gamma_{5}d_{b}-d_{a}^{T}C\gamma_{5}u_{b})Q_{c},\nonumber \\
J_{\Xi_{Q}} & = & \frac{1}{\sqrt{2}}\epsilon_{abc}(q_{a}^{T}C\gamma_{5}s_{b}-s_{a}^{T}C\gamma_{5}q_{b})Q_{c},\label{eq:current_anti_triplet}
\end{eqnarray}
and for the SU(3) sextet they are
\begin{eqnarray}
J_{\Sigma_{Q}} & = & \frac{1}{\sqrt{2}}\epsilon_{abc}(u_{a}^{T}C\gamma^{\mu}d_{b}+d_{a}^{T}C\gamma^{\mu}u_{b})\gamma_{\mu}\gamma_{5}Q_{c},\nonumber \\
J_{\Xi_{Q}^{\prime}} & = & \frac{1}{\sqrt{2}}\epsilon_{abc}(q_{a}^{T}C\gamma^{\mu}s_{b}+s_{a}^{T}C\gamma^{\mu}q_{b})\gamma_{\mu}\gamma_{5}Q_{c},\nonumber \\
J_{\Omega_{Q}} & = & \epsilon_{abc}s_{a}^{T}C\gamma^{\mu}s_{b}\gamma_{\mu}\gamma_{5}Q_{c}.\label{eq:current_sextet}
\end{eqnarray}
Similar definitions for the interpolating currents were adopted in Refs. \cite{Zhang:2008rt,Wang:2010hs,Wang:2010fq}, and some discussions can be found in Ref. \cite{Zhang:2008rt}.

The correlation function can be calculated at both hadron and QCD
level. In the following, we will only present the extraction
of the vector-current form factors, and the axial-vector-current form
factors can be determined in a similar way. At hadron level, one can
insert complete sets of the initial and final hadronic states into
the correlation function and consider the contributions from positive and negative parity baryons
simultaneously, then the correlation function can be written as
\begin{eqnarray}
\Pi_{\mu}^{V,{\rm had}} & = & \lambda_{f}^{+}\lambda_{i}^{+}\frac{(\slashed p_{2}+M_{2}^{+})(\frac{p_{1\mu}}{M_{1}^{+}}F_{1}^{++}+\frac{p_{2\mu}}{M_{2}^{+}}F_{2}^{++}+\gamma_{\mu}F_{3}^{++})(\slashed p_{1}+M_{1}^{+})}{(p_{2}^{2}-M_{2}^{+2})(p_{1}^{2}-M_{1}^{+2})}\nonumber \\
 & + & \lambda_{f}^{+}\lambda_{i}^{-}\frac{(\slashed p_{2}+M_{2}^{+})(\frac{p_{1\mu}}{M_{1}^{-}}F_{1}^{+-}+\frac{p_{2\mu}}{M_{2}^{+}}F_{2}^{+-}+\gamma_{\mu}F_{3}^{+-})(\slashed p_{1}-M_{1}^{-})}{(p_{2}^{2}-M_{2}^{+2})(p_{1}^{2}-M_{1}^{-2})}\nonumber \\
 & + & \lambda_{f}^{-}\lambda_{i}^{+}\frac{(\slashed p_{2}-M_{2}^{-})(\frac{p_{1\mu}}{M_{1}^{+}}F_{1}^{-+}+\frac{p_{2\mu}}{M_{2}^{-}}F_{2}^{-+}+\gamma_{\mu}F_{3}^{-+})(\slashed p_{1}+M_{1}^{+})}{(p_{2}^{2}-M_{2}^{-2})(p_{1}^{2}-M_{1}^{+2})}\nonumber \\
 & + & \lambda_{f}^{-}\lambda_{i}^{-}\frac{(\slashed p_{2}-M_{2}^{-})(\frac{p_{1\mu}}{M_{1}^{-}}F_{1}^{--}+\frac{p_{2\mu}}{M_{2}^{-}}F_{2}^{--}+\gamma_{\mu}F_{3}^{--})(\slashed p_{1}-M_{1}^{-})}{(p_{2}^{2}-M_{2}^{-2})(p_{1}^{2}-M_{1}^{-2})}\nonumber \\
 & + & \cdots.\label{eq:correlator_hadronic}
\end{eqnarray}
In Eq. (\ref{eq:correlator_hadronic}), the ellipsis stands for the
contribution from higher resonances and continuum spectra, $M_{1(2)}^{+(-)}$
denotes the mass of the initial (final) positive (negative) parity
baryons, and $F_{1}^{-+}$ is the form factor $F_{1}$ defined
in Eq. (\ref{eq:parameterization_2}) with the negative-parity final state and the positive-parity initial state, and so
forth. To arrive at Eq. (\ref{eq:correlator_hadronic}), we have
adopted the pole residue definitions for positive and negative parity
baryons
\begin{eqnarray}
\langle0|J_{+}|{\cal B}_{+}(p,s)\rangle & = & \lambda_{+}u(p,s),\nonumber \\
\langle0|J_{+}|{\cal B}_{-}(p,s)\rangle & = & (i\gamma_{5})\lambda_{-}u(p,s),\label{eq:pole_residue}
\end{eqnarray}
and the following conventions for the form factors $F_{i}^{\pm\pm}$:
\begin{eqnarray}
\langle{\cal B}_{f}^{+}(p_{2},s_{2})|V_{\mu}|{\cal B}_{i}^{+}(p_{1},s_{1})\rangle & = & \bar{u}(p_{2},s_{2})[\frac{p_{1\mu}}{M_{1}^{+}}F_{1}^{++}+\frac{p_{2\mu}}{M_{2}^{+}}F_{2}^{++}+\gamma_{\mu}F_{3}^{++}]u(p_{1},s_{1}),\nonumber \\
\langle{\cal B}_{f}^{+}(p_{2},s_{2})|V_{\mu}|{\cal B}_{i}^{-}(p_{1},s_{1})\rangle & = & \bar{u}(p_{2},s_{2})[\frac{p_{1\mu}}{M_{1}^{-}}F_{1}^{+-}+\frac{p_{2\mu}}{M_{2}^{+}}F_{2}^{+-}+\gamma_{\mu}F_{3}^{+-}](i\gamma_{5})u(p_{1},s_{1}),\nonumber \\
\langle{\cal B}_{f}^{-}(p_{2},s_{2})|V_{\mu}|{\cal B}_{i}^{+}(p_{1},s_{1})\rangle & = & \bar{u}(p_{2},s_{2})(i\gamma_{5})[\frac{p_{1\mu}}{M_{1}^{+}}F_{1}^{-+}+\frac{p_{2\mu}}{M_{2}^{-}}F_{2}^{-+}+\gamma_{\mu}F_{3}^{-+}]u(p_{1},s_{1}),\nonumber \\
\langle{\cal B}_{f}^{-}(p_{2},s_{2})|V_{\mu}|{\cal B}_{i}^{-}(p_{1},s_{1})\rangle & = & \bar{u}(p_{2},s_{2})(i\gamma_{5})[\frac{p_{1\mu}}{M_{1}^{-}}F_{1}^{--}+\frac{p_{2\mu}}{M_{2}^{-}}F_{2}^{--}+\gamma_{\mu}F_{3}^{--}](i\gamma_{5})u(p_{1},s_{1}).
\end{eqnarray}
In Eq. (\ref{eq:pole_residue}), $J_{+}$ can be found in Eqs. (\ref{eq:current_QQ}-\ref{eq:current_sextet}), and  $\lambda_{+(-)}$ is the pole residue for the positive (negative)
parity baryon.

At the QCD level, the correlation function can be evaluated using
the operator product expansion (OPE), and expanded as a power of matrix
elements of local operators in the deep Euclidean momentum region.
This expansion is organized by the inverse of mass dimensions. The
identity operator corresponds to the so-called perturbative term and
higher dimensional operators are called the condensate terms. A detailed
calculation of these contributions will be presented in the following
subsections, including the perturbative contribution (dim-0), the
quark condensate contribution (dim-3) and the mixed quark-gluon condensate
contribution (dim-5). For practical use, it is convenient to express
the correlation function as a double dispersion relation 
\begin{equation}
\Pi_{\mu}^{V,{\rm QCD}}(p_{1}^{2},p_{2}^{2},q^{2})=\int^{\infty}ds_{1}\int^{\infty}ds_{2}\frac{\rho_{\mu}^{V,{\rm QCD}}(s_{1},s_{2},q^{2})}{(s_{1}-p_{1}^{2})(s_{2}-p_{2}^{2})},\label{eq:correlator_QCD}
\end{equation}
with $\rho_{\mu}^{V,{\rm QCD}}(s_{1},s_{2},q^{2})$ being the spectral
function, which can be obtained by applying Cutkosky cutting rules.
Quark-hadron duality guarantees that results for correlation functions
derived at hadron level and QCD level are equivalent. In particular,
it is plausible to identify the spectral functions above threshold
at the hadron level and QCD level. Under this assumption, the sum of the four pole terms
in Eq. (\ref{eq:correlator_hadronic}) should be equal to
\begin{equation}
\int^{s_{1}^{0}}ds_{1}\int^{s_{2}^{0}}ds_{2}\frac{\rho_{\mu}^{V,{\rm QCD}}(s_{1},s_{2},q^{2})}{(s_{1}-p_{1}^{2})(s_{2}-p_{2}^{2})}\equiv\Pi_{\mu}^{V,{\rm pole}},
\end{equation}
where $s_{1(2)}^{0}$ is the threshold parameter for the initial
(final) baryon.
$\Pi_{\mu}^{V,{\rm pole}}$ can be formally written as
\begin{equation}
\Pi_{\mu}^{V,{\rm pole}}=\sum_{i=1}^{12}A_{i}e_{i\mu},\label{eq:correlator_pole_formal}
\end{equation}
where, for latter convenience, we define
\begin{eqnarray}
(e_{1,2,3,4})_{\mu} & = & \{\slashed p_{2},1\}\times\{p_{1\mu}\}\times\{\slashed p_{1},1\},\nonumber \\
(e_{5,6,7,8})_{\mu} & = & \{\slashed p_{2},1\}\times\{p_{2\mu}\}\times\{\slashed p_{1},1\},\nonumber \\
(e_{9,10,11,12})_{\mu} & = & \{\slashed p_{2},1\}\times\{\gamma_{\mu}\}\times\{\slashed p_{1},1\}.
\label{eq:e_i_mu}
\end{eqnarray}
Then one can obtain these 12 form factors $F_{i}^{\pm,\pm}$ in Eq. \eqref{eq:correlator_hadronic} by comparing the corresponding coefficients of these 12 Dirac structures at hadronic and QCD levels. Especially, one can obtain the expressions
for $F_{i}^{++}$ as:

\begin{eqnarray}
\frac{\lambda_{i}^{+}\lambda_{f}^{+}(F_{1}^{++}/M_{1}^{+})}{(p_{1}^{2}-M_{1}^{+2})(p_{2}^{2}-M_{2}^{+2})} & = & \frac{\{M_{1}^{-}M_{2}^{-},M_{2}^{-},M_{1}^{-},1\}.\{A_{1},A_{2},A_{3},A_{4}\}}{(M_{1}^{+}+M_{1}^{-})(M_{2}^{+}+M_{2}^{-})},\nonumber \\
\frac{\lambda_{i}^{+}\lambda_{f}^{+}(F_{2}^{++}/M_{2}^{+})}{(p_{1}^{2}-M_{1}^{+2})(p_{2}^{2}-M_{2}^{+2})} & = & \frac{\{M_{1}^{-}M_{2}^{-},M_{2}^{-},M_{1}^{-},1\}.\{A_{5},A_{6},A_{7},A_{8}\}}{(M_{1}^{+}+M_{1}^{-})(M_{2}^{+}+M_{2}^{-})},\nonumber \\
\frac{\lambda_{i}^{+}\lambda_{f}^{+}F_{3}^{++}}{(p_{1}^{2}-M_{1}^{+2})(p_{2}^{2}-M_{2}^{+2})} & = & \frac{\{M_{1}^{-}M_{2}^{-},M_{2}^{-},M_{1}^{-},1\}.\{A_{9},A_{10},A_{11},A_{12}\}}{(M_{1}^{+}+M_{1}^{-})(M_{2}^{+}+M_{2}^{-})}.
\end{eqnarray}
In practice Borel transformation are usually adopted to improve the
convergence in the quark-hadron duality and suppress the higher resonance
and continuum contributions:
\begin{align}
\lambda_{i}^{+}\lambda_{f}^{+}(F_{1}^{++}/M_{1}^{+})\exp\left(-\frac{M_{1}^{+2}}{T_{1}^{2}}-\frac{M_{2}^{+2}}{T_{2}^{2}}\right) & =\frac{\{M_{1}^{-}M_{2}^{-},M_{2}^{-},M_{1}^{-},1\}.\{{\cal B}A_{1},{\cal B}A_{2},{\cal B}A_{3},{\cal B}A_{4}\}}{(M_{1}^{+}+M_{1}^{-})(M_{2}^{+}+M_{2}^{-})},\nonumber \\
\lambda_{i}^{+}\lambda_{f}^{+}(F_{2}^{++}/M_{2}^{+})\exp\left(-\frac{M_{1}^{+2}}{T_{1}^{2}}-\frac{M_{2}^{+2}}{T_{2}^{2}}\right) & =\frac{\{M_{1}^{-}M_{2}^{-},M_{2}^{-},M_{1}^{-},1\}.\{{\cal B}A_{5},{\cal B}A_{6},{\cal B}A_{7},{\cal B}A_{8}\}}{(M_{1}^{+}+M_{1}^{-})(M_{2}^{+}+M_{2}^{-})},\nonumber \\
\lambda_{i}^{+}\lambda_{f}^{+}F_{3}^{++}\exp\left(-\frac{M_{1}^{+2}}{T_{1}^{2}}-\frac{M_{2}^{+2}}{T_{2}^{2}}\right) & =\frac{\{M_{1}^{-}M_{2}^{-},M_{2}^{-},M_{1}^{-},1\}.\{{\cal B}A_{9},{\cal B}A_{10},{\cal B}A_{11},{\cal B}A_{12}\}}{(M_{1}^{+}+M_{1}^{-})(M_{2}^{+}+M_{2}^{-})},
\label{eq:Fi_plus_plus}
\end{align}
where ${\cal B}A_{i}\equiv{\cal B}_{T_{1}^{2},T_{2}^{2}}A_{i}$ are
doubly Borel transformed coefficients, and $T_{1}^{2}$ and $T_{2}^{2}$
are the Borel mass parameters.

The coefficients $A_{i}$ in Eq. (\ref{eq:correlator_pole_formal})
can be projected out in the following way. Multiplying by $e_{j}^{\mu}$
then taking the trace on the both sides of Eq. (\ref{eq:correlator_pole_formal}),
one can arrive at the following 12 linear equations: 
\begin{equation}
B_{j}\equiv{\rm Tr}[\Pi_{\mu}^{V,{\rm pole}}e_{j}^{\mu}]={\rm Tr}\left[\left(\sum_{i=1}^{12}A_{i}e_{i\mu}\right)e_{j}^{\mu}\right],\quad j=1,...,12,
\end{equation}
Solving these equations one can obtain these $A_{i}$.

In the following, we will use the vector-current form factors for
doubly-heavy baryon into a SU(3) sextet baryon as an example to illustrate
our calculation. Results for other transitions can be obtained in
a similar manner.
	
	\subsection{The perturbative contribution}

\begin{figure}
\includegraphics[width=0.35\columnwidth]{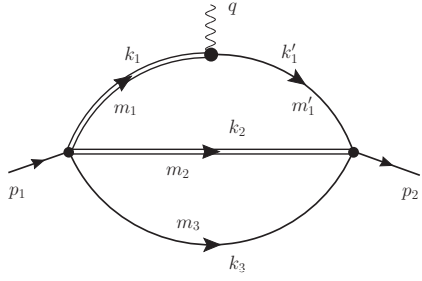} \caption{The perturbative contribution to transition form factors. The doubly-solid line denotes a heavy quark, and the ordinary solid line corresponds to a light quark. }
\label{fig:pert} 
\end{figure}

The perturbative contribution is derived by computing  the coefficient of the identity operator in   OPE. The corresponding   Feynman diagram is shown in Fig.~\ref{fig:pert}.  The doubly-solid line denotes a heavy bottom/charm quark, and the ordinary solid line corresponds to a light quark. Its contribution   is given as
\begin{eqnarray}
	\Pi_{\mu}^{V,{\rm pert}}(p_{1}^{2},p_{2}^{2},q^{2}) & = & 6\cdot 2\sqrt{2}\ i^{2}\int\frac{d^{4}k_{2}}{(2\pi)^{4}}\frac{d^{4}k_{3}}{(2\pi)^{4}}\frac{N_{\mu}}{(k_{1}^{2}-m_{1}^{2})(k_{1}^{\prime2}-m_{1}^{\prime2})(k_{2}^{2}-m_{2}^{2})(k_{3}^{2}-m_{3}^{2})},\label{eq;correlator_pert}
	\end{eqnarray}
where the factor 6 comes from the color   contraction $\epsilon_{abc}\epsilon^{abc}$, the factor $2\sqrt{2}$ comes from the contraction   of quark fields and the normalization factors of the baryon currents. The numerator of the integrand in Eq.~(\ref{eq;correlator_pert}) is:
	\begin{eqnarray}
	N_{\mu} & = & \gamma_{\alpha^{\prime}}\gamma_{5}(\slashed k_{2}+m_{2})\gamma^{\alpha}(\slashed k_{1}-m_{1})\gamma_{\mu}(\slashed k_{1}^{\prime}-m_{1}^{\prime})\gamma^{\alpha^{\prime}}(\slashed k_{3}+m_{3})\gamma_{\alpha}\gamma_{5},\nonumber \\
	k_{1} & = & p_{1}-k_{2}-k_{3},\ \ \ k_{1}^{\prime} = p_{2}-k_{2}-k_{3}. 
	\end{eqnarray}
The correlation function can be expressed in terms of a double dispersion integration: 
	\begin{equation}
	\Pi_{\mu}^{V,{\rm pert}}(p_{1}^{2},p_{2}^{2},q^{2})=\int ds_{1}ds_{2}\frac{\rho_{\mu}^{V,{\rm pert}}(s_{1},s_{2},q^{2})}{(s_{1}-p_{1}^{2})(s_{2}-p_{2}^{2})}.
\end{equation}
Here the spectral function $\rho_{\mu}^{V,{\rm pert}}(s_{1},s_{2},q^{2})$ is proportional to the discontinuity of the correlation function with respect to $s_{1}$ and $s_{2}$. According to the Cutkosky rule, the spectral function can be obtained by setting all the propagators onshell: 
\begin{equation}
	\rho_{\mu}^{V,{\rm pert}}(s_{1},s_{2},q^{2})=\frac{(-2\pi i)^{4}}{(2\pi i)^{2}}(12\sqrt{2}i^{2})\int\frac{d^{4}k_{2}}{(2\pi)^{4}}\frac{d^{4}k_{3}}{(2\pi)^{4}}\delta(k_{1}^{2}-m_{1}^{2})\delta(k_{1}^{\prime2}-m_{1}^{\prime2})\delta(k_{2}^{2}-m_{2}^{2})\delta(k_{3}^{2}-m_{3}^{2})N_{\mu}.\label{eq:rho_pert}
	\end{equation}
The phase-space-like integral can be evaluated as:
	\begin{equation}
	\int d^{4}k_{2}d^{4}k_{3}\delta(k_{1}^{2}-m_{1}^{2})\delta(k_{1}^{\prime2}-m_{1}^{\prime2})\delta(k_{2}^{2}-m_{2}^{2})\delta(k_{3}^{2}-m_{3}^{2})=\int dm_{23}^{2}\int_{\triangle}\int_{2},
	\end{equation}
	where 
	\begin{eqnarray}
	\int_{\triangle} & \equiv & \int d^{4}k_{1}d^{4}k_{1}^{\prime}d^{4}k_{23}\delta(k_{1}^{2}-m_{1}^{2})\delta(k_{1}^{\prime2}-m_{1}^{\prime2})\delta(k_{23}^{2}-m_{23}^{2})\delta^{4}(p_{1}-k_{1}-k_{23})\delta^{4}(p_{2}-k_{1}^{\prime}-k_{23}),\label{eq:triangle_phase_space}\nonumber\\
\int_{2} & \equiv & \int d^{4}k_{2}d^{4}k_{3}\delta(k_{2}^{2}-m_{2}^{2})\delta(k_{3}^{2}-m_{3}^{2})\delta^{4}(k_{23}-k_{2}-k_{3}). \label{eq:two_body_phase_space} 
\end{eqnarray}

\subsection{The quark condensate contribution}

\begin{figure}
\includegraphics[width=0.75\columnwidth]{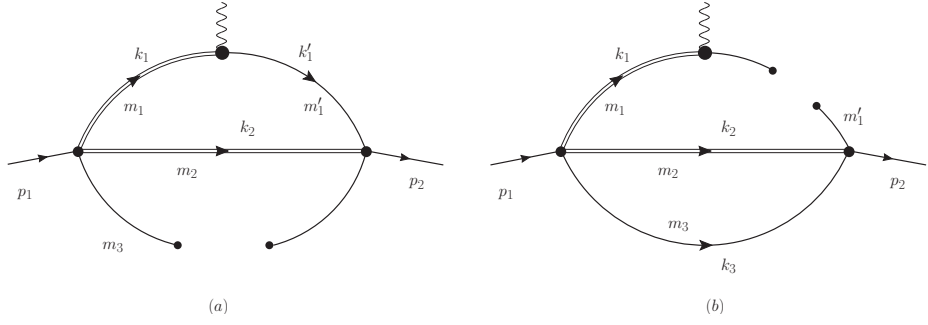} \caption{Light-quark condensate diagrams. Heavy quark will not condensate and thus only the two light-quark propagators give condensate contributions.   }
\label{fig:qqCondense} 
\end{figure}

The $\bar q q$ condensate operator in the OPE has dimension 3, and its Feynman diagram is shown in Fig.~\ref{fig:qqCondense}. Since heavy quarks will not contribute with condensations,  there are two diagrams from the light quark condensate. 
The diagram (\ref{fig:qqCondense}a) gives: 
\begin{equation}
	\Pi_{\mu}^{V,\langle\bar{q}q\rangle,a}(p_{1}^{2},p_{2}^{2},q^{2})=(-6\cdot2\sqrt{2}i)\frac{1}{12}\langle\bar{q}q\rangle\int\frac{d^{4}k_{2}}{(2\pi)^{4}}\frac{N_{\mu}^{V,\langle\bar{q}q\rangle,a}}{(k_{1}^{2}-m_{1}^{2})(k_{1}^{\prime2}-m_{1}^{\prime2})(k_{2}^{2}-m_{2}^{2})},
\end{equation}
where the condensate term is defined as $\langle q_{a}^{i}\bar{q}_{b}^{j}\rangle= - (\langle\bar{q}q\rangle/12)\delta_{ab}\delta^{ij}$, and the numerator is:
	\begin{eqnarray}
		N_{\mu}^{V,\langle\bar{q}q\rangle,a} & = & \gamma_{\alpha^{\prime}}\gamma_{5}(\slashed k_{2}+m_{2})\gamma^{\alpha}(\slashed k_{1}-m_{1})\gamma_{\mu}(\slashed k_{1}^{\prime}-m_{1}^{\prime})\gamma^{\alpha^{\prime}}\gamma_{\alpha}\gamma_{5},\nonumber\\
		k_{1} & = & p_{1}-k_{2},\ \ \ k_{1}^{\prime} = p_{2}-k_{2}.
	\end{eqnarray}

According to the Cutkosky rule, the   spectral function can now be evaluated  as: 
	\begin{equation}
	\rho_{\mu}^{V,\langle\bar{q}q\rangle,a}(s_{1},s_{2},q^{2})=\frac{1}{(2\pi i)^{2}}(-2\pi i)^{3}(-\sqrt{2}i)\langle\bar{q}q\rangle\frac{1}{(2\pi)^{4}}\int_{\triangle}N_{\mu}^{V,\langle\bar{q}q\rangle,a},\label{eq:rho_qbarq}
	\end{equation}
where the integral $\int_{\triangle}$ is slightly different from that in Eq.~(\ref{eq:triangle_phase_space}),  with  $m_{23}^{2}$ being replaced by $m_{2}^{2}$. The   diagram (b) has the amplitude: 
\begin{equation}
	\Pi_{\mu}^{V,\langle\bar{q}q\rangle,b}(p_{1}^{2},p_{2}^{2},q^{2})\sim\int\frac{d^{4}k_{2}}{(2\pi)^{4}}\frac{N_{\mu}^{V,\langle\bar{q}q\rangle,b}}{(q^{2}-m_{1}^{2})(k_{2}^{2}-m_{2}^{2})((p_{2}-k_{2})^{2}-m_{3}^{2})}.
	\end{equation}
One can see that the denominator is independent of $p_{1}^{2}$, and  thereby the corresponding double discontinuity must vanish. As a result, the quark condensate contribution only comes from  Fig.~(\ref{fig:qqCondense}a).
	
\subsection{Mixed quark-gluon condensate contribution}

\begin{figure}
\includegraphics[width=0.6\columnwidth]{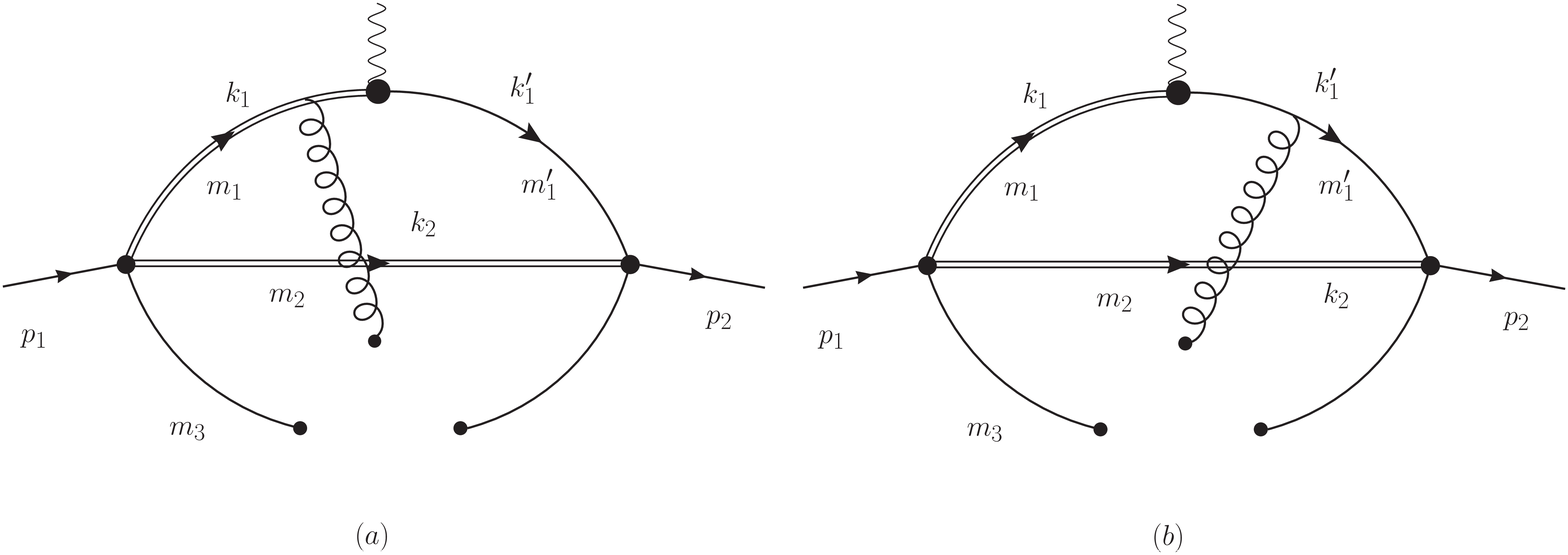} \includegraphics[width=0.3\columnwidth]{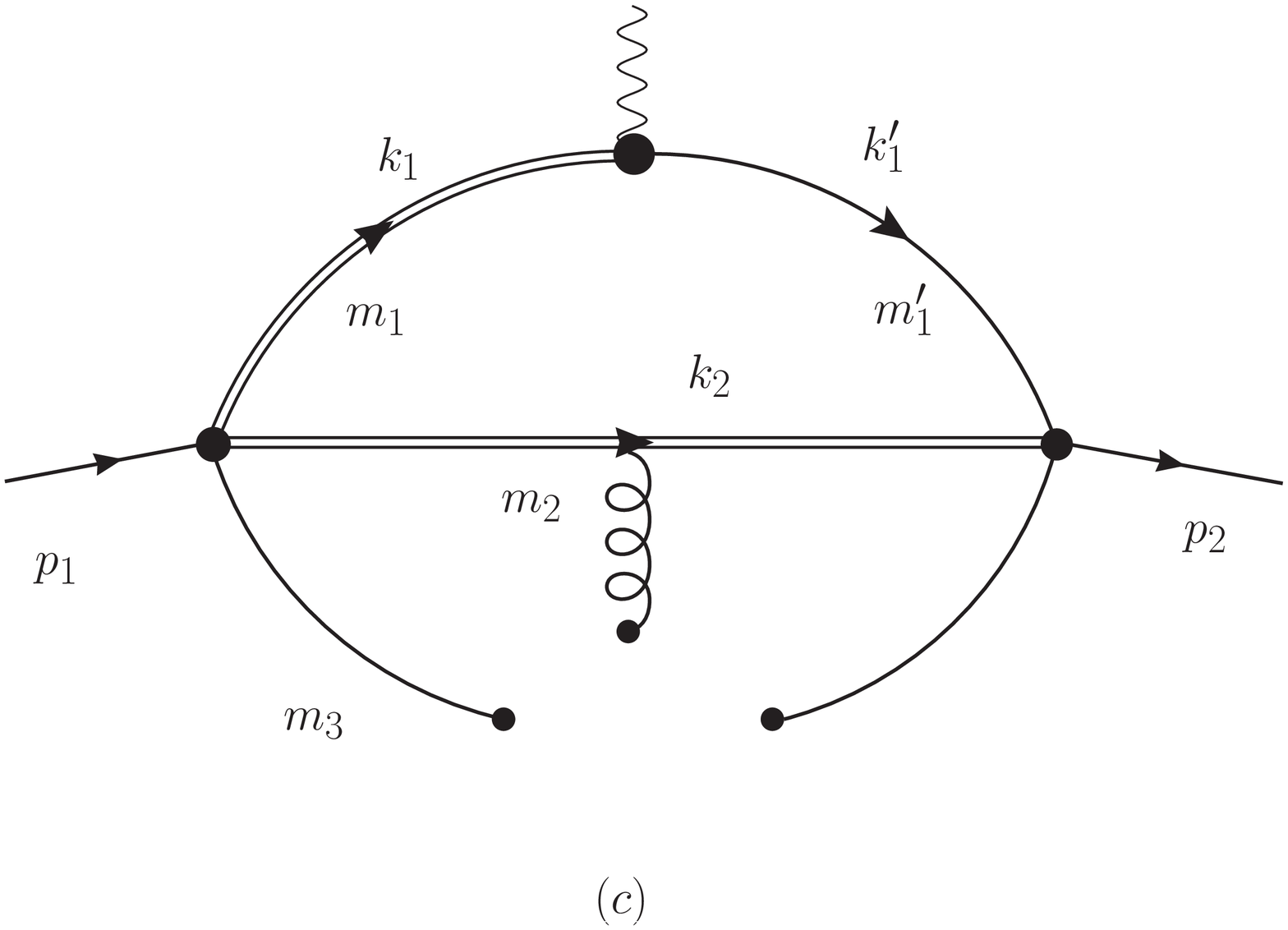}
\caption{Mixed quark-gluon condensate diagrams.}
\label{fig:qGqcondensate} 
\end{figure}

\begin{figure}
\includegraphics[width=0.8\columnwidth]{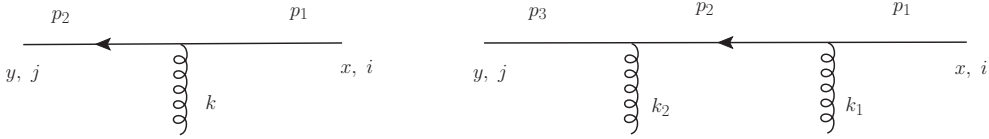} \caption{Quark propagators in the QCD vaccum. $x$ and $y$ are spacetime coordinates, $i$ and $j$ are color indices, and $p_{i}$, $k$ and $k_{i}$ are momenta. }
\label{fig:propagators} 
\end{figure}

The quark-gluon condensate operator  $\bar q g_s G q$ has dimension 5   in OPE. There are three Feynman diagrams for mixed quark-gluon condensate contribution, as shown in Fig.~\ref{fig:qGqcondensate}.  We are requested
to consider the interaction of the propagating quark with the background gluons. The quark propagators with one gluon and two gluons attached (Fig.~\ref{fig:propagators}) respectively have the following forms: 
	\begin{eqnarray}
	S^{(1)ji}(x,\ y) & = & ig\int\frac{d^{4}p_{2}}{(2\pi)^{4}}\int\frac{d^{4}k}{(2\pi)^{4}}e^{-ip_{2}\cdot y}e^{i(p_{2}-k)\cdot x}\tilde{A}_{\mu}^{ji}(k)\frac{i}{\slashed p_{2}-m}\gamma^{\mu}\frac{i}{\slashed p_{2}-\slashed k-m},\nonumber \\
	S^{(2)ji}(x,\ y) & = & (ig)^{2}\int\frac{d^{4}p_{3}}{(2\pi)^{4}}\int\frac{d^{4}k_{2}}{(2\pi)^{4}}\int\frac{d^{4}k_{1}}{(2\pi)^{4}}e^{-ip_{3}\cdot y}e^{i(p_{3}-k_{2}-k_{1})\cdot x}(\tilde{A}_{\nu}(k_{2})\tilde{A}_{\mu}(k_{1}))^{ji}\nonumber \\
	&  & \times\frac{i}{\slashed p_{3}-m}\gamma^{\nu}\frac{i}{\slashed p_{3}-\slashed k_{2}-m}\gamma^{\mu}\frac{i}{\slashed p_{3}-\slashed k_{2}-\slashed k_{1}-m}.
	\end{eqnarray}
	In the fixed-point gauge, the background gluon field expanded to the
	lowest order (in the momentum space) is: 
	\begin{equation}
	\tilde{A}_{\mu}^{a}(k)=-\frac{i}{2}(2\pi)^{4}G_{\alpha\mu}^{a}(0)\frac{\partial}{\partial k_{\alpha}}\delta^{4}(k).
	\end{equation}
Thus a propagating quark can exchange arbitrary  numbers of zero momentum gluons with the QCD vacuum. It should be noted that the fixed-point gauge violates the spacetime	translation invariance. As a result, $S(x,y)$ is not the same as $S(x-y,0)$.
In the cases of quark-gluon condensate contribution as well as gluon-gluon condensate contribution to be discussed in the following, the following formulas are useful: 
\begin{eqnarray*}
\int d^{4}uf(u)\frac{\partial}{\partial u_{\alpha}}\delta^{4}(u) & = & -\frac{\partial}{\partial u_{\alpha}}f(u)\Big|_{u=0},\\
\frac{\partial}{\partial u_{\alpha}}\frac{1}{\slashed p+\slashed u-m}\Big|_{u=0} & = & -\frac{1}{\slashed p-m}\gamma^{\alpha}\frac{1}{\slashed p-m},
\end{eqnarray*}
where $u$ stands for the momentum of the soft gluon, and $f(u)$
is an arbitrary function of $u$.

In Fig~(\ref{fig:qGqcondensate}a), the upper left heavy quark interacts with a background gluon field, which condensates with the two light quark fields. Its contribution is given as: 
	\begin{equation}
	\Pi_{\mu}^{V,\langle\bar{q}Gq\rangle,a}(p_{1}^{2},p_{2}^{2},q^{2})=-\frac{\sqrt{2}}{192}{\rm Tr}[T^{a}T^{a}]\langle\bar{q}g_{s}\sigma Gq\rangle\int\frac{d^{4}k_{2}}{(2\pi)^{4}}\frac{N_{\mu}^{V,\langle\bar{q}Gq\rangle,a}}{(k_{1}^{2}-m_{1}^{2})^{3}(k_{1}^{\prime2}-m_{1}^{\prime2})(k_{2}^{2}-m_{2}^{2})}. 
	\label{eq:correlator_qGq_a}
	\end{equation}
The condensate term is defined as $\langle q^i_a g_s G^c_{\mu\nu} \bar q^j_b\rangle = -(1/192)\langle\bar{q} g_s \sigma G q\rangle (\sigma_{\mu\nu})^{ij}T^c_{ab}$, and the numerator is:
	\begin{eqnarray}
		N_{\mu}^{V,\langle\bar{q}Gq\rangle,a} & = & \gamma_{\nu^{\prime}}\gamma_{5}(\slashed k_{2}+m_{2})\gamma^{\nu}(\slashed k_{1}-m_{1})\gamma^{\alpha}(\slashed k_{1}-m_{1})\gamma^{\rho}(\slashed k_{1}-m_{1})\gamma_{\mu}(\slashed k_{1}^{\prime}-m_{1}^{\prime})\gamma^{\nu^{\prime}}\sigma_{\rho\alpha}\gamma_{\nu}\gamma_{5},\nonumber\\
		k_{1} & = & p_{1}-k_{2},\ \ \ k_{1}^{\prime} = p_{2}-k_{2}.
	\end{eqnarray}
In Eq. \eqref{eq:correlator_qGq_a}, $1/(k_{1}^{2}-m_{1}^{2})^{3}$ can be  handled  in a derivative method: 
	\begin{equation}
	\frac{1}{(k_{1}^{2}-m_{1}^{2})^{n}}=\frac{1}{(n-1)!}\frac{\partial^{n-1}}{(\partial m_{1s})^{n-1}}\left(\frac{1}{k_{1}^{2}-m_{1s}}\right)\Bigg|_{m_{1s}=m_{1}^{2}}.
	\end{equation}
Then the spectral function can be derived by using  Cutkosky rule before applying  the mass derivative: 
\begin{eqnarray}
\rho_{\mu}^{V,\langle\bar{q}Gq\rangle,a}(p_{1}^{2},p_{2}^{2},q^{2}) & = & \frac{(-2\pi i)^{3}}{(2\pi i)^{2}}(-\frac{\sqrt{2}}{192}){\rm Tr}[T^{a}T^{a}]\langle\bar{q}g_{s}\sigma Gq\rangle\frac{1}{(2\pi)^{4}}\nonumber \\
 &  & \times\left(\frac{1}{2}\frac{\partial^{2}}{(\partial m_{1s})^{2}}\int_{\triangle}N_{\mu}^{V,\langle\bar{q}Gq\rangle,a}\Big|_{k_{1}^{2}\to m_{1s}}\right)\Big|_{m_{1s}\to m_{1}^{2}},
\end{eqnarray}
The  the integral $\int_{\triangle}$ is slightly different from
that in Eq.~(\ref{eq:rho_qbarq}),  with  $m_{1}^{2}$ being replaced by $m_{1s}$. The other two diagrams in Fig.~\ref{fig:qGqcondensate} can be calculated similarly. 

\subsection{Gluon-gluon condensate contribution}

\begin{figure}
\includegraphics[width=0.35\columnwidth]{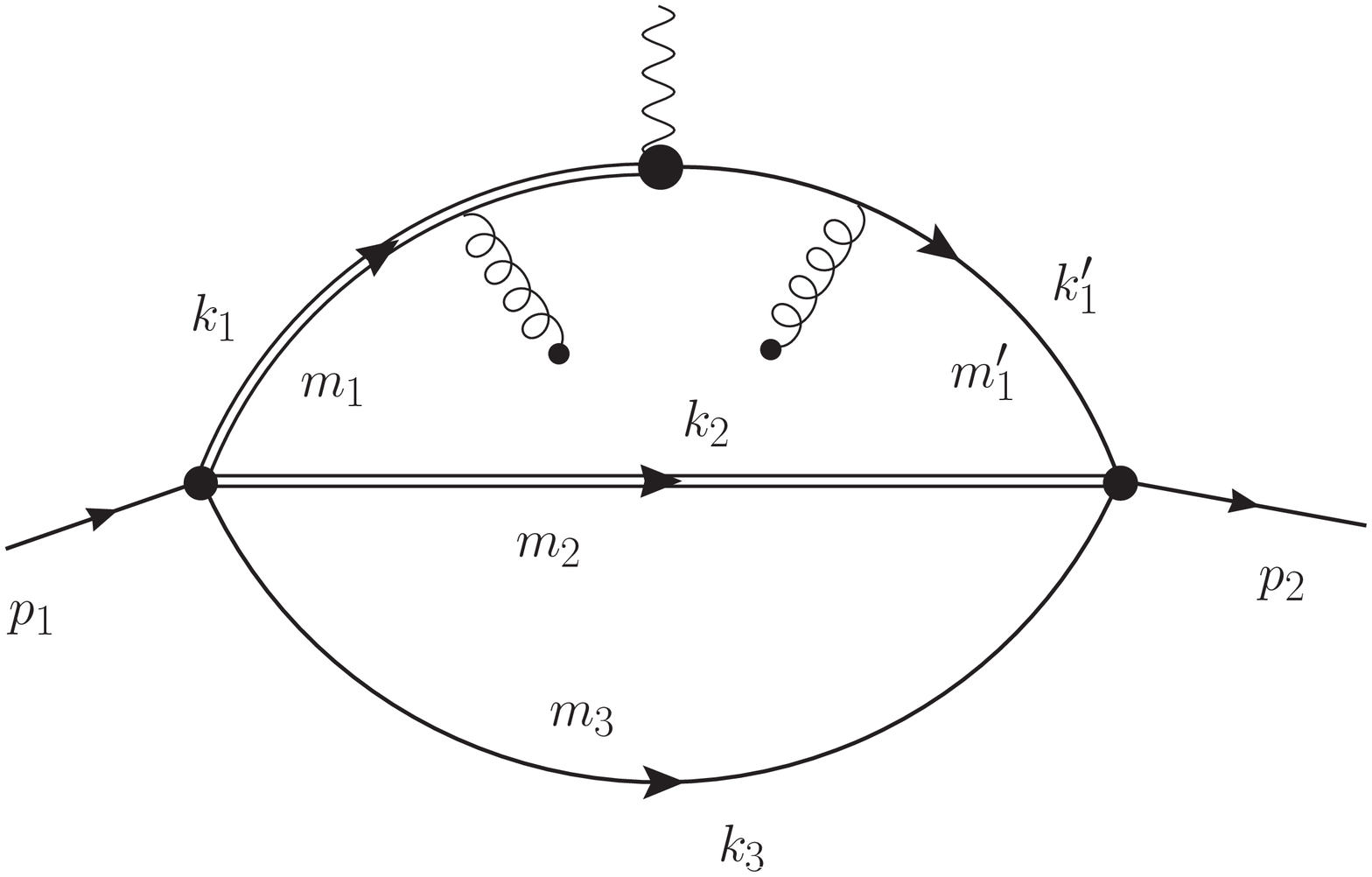} \caption{One of the gluon-gluon condensate diagrams. }
\label{fig:GG2} 
\end{figure}

In the case of the dim-4 operator $GG$ in the OPE, i.e. the gluon-gluon
condensate, two background gluon fields interact with the four quark
propagators, and one example is shown in Fig.~\ref{fig:GG2}.

The contribution from Fig.~\ref{fig:GG2} is: 
\begin{eqnarray*}
\Pi_{\mu}^{V,\langle GG\rangle}(p_{1}^{2},p_{2}^{2},q^{2}) & = & \frac{\langle g_{s}^{2}GG\rangle}{48\sqrt{2}}{\rm Tr}[T^{a}T^{a}]\int\frac{d^{4}k_{2}}{(2\pi)^{4}}\frac{d^{4}k_{3}}{(2\pi)^{4}}(-g_{\alpha\sigma}g_{\rho\beta}+g_{\alpha\beta}g_{\rho\sigma})\\
 &  & \times\Big(-\gamma_{\nu^{\prime}}\gamma_{5}\frac{1}{\slashed k_{2}-m_{2}}\gamma^{\nu}\frac{1}{\slashed k_{1}+m_{1}}\gamma^{\alpha}\frac{1}{\slashed k_{1}+m_{1}}\gamma^{\rho}\frac{1}{\slashed k_{1}+m_{1}}\\
 &  & \times\gamma_{\mu}\frac{1}{\slashed k_{1}^{\prime}+m_{1}^{\prime}}\gamma^{\sigma}\frac{1}{\slashed k_{1}^{\prime}+m_{1}^{\prime}}\gamma^{\beta}\frac{1}{\slashed k_{1}^{\prime}+m_{1}^{\prime}}\gamma^{\nu^{\prime}}\frac{1}{\slashed k_{3}-m_{3}}\gamma_{\nu}\gamma_{5}\Big).
\end{eqnarray*}
Note that $\Pi_{\mu}^{V,\langle GG\rangle}(p_{1}^{2},p_{2}^{2},q^{2})$
contains 19 Dirac matrices.

Similar procedure can be applied to extract the spectral function, and the corresponding numerical results will be shown
in Sec.~\ref{sec:numerical}.

\section{Numerical results}
\label{sec:numerical}

The input parameters used in our numerical calculation  are taken as~\cite{Ioffe:2005ym,Colangelo:2000dp,Olive:2016xmw,Tanabashi:2018oca}:
$\langle\bar{q}q\rangle=-(0.24\pm0.01{\rm GeV})^{3}$, $\langle\bar{s}s\rangle=(0.8\pm0.2)\langle\bar{q}q\rangle$,
$\langle\bar{q}g_{s}\sigma Gq\rangle=m_{0}^{2}\langle\bar{q}q\rangle$,
$\langle\bar{s}g_{s}\sigma Gs\rangle=m_{0}^{2}\langle\bar{s}s\rangle$,
$m_{0}^{2}=(0.8\pm0.2)\ {\rm GeV}^{2}$, $\langle\frac{\alpha_{s}GG}{\pi}\rangle=(0.012\pm0.004)\ {\rm GeV}^{4}$
for the condensate parameters and $m_{s}=(0.14\pm0.01)\ {\rm GeV}$, $m_{c}=(1.35\pm0.10)\ {\rm GeV}$, $m_{b}=(4.7\pm0.1)\ {\rm GeV}$ for the quark masses. The pole residues of doubly-heavy and singly-heavy baryons as well
as their masses are collected in Table~\ref{Tab:decay_constant}.
The factor $\sqrt{2}$ in Table~\ref{Tab:decay_constant} arises
from the convention difference in the definitions of the interpolating
current for baryon~\cite{Wang:2010fq,Wang:2009cr,Hu:2017dzi}. For
doubly-heavy baryons, we have updated the pole residues using the
same inputs as those in this work. The mass of $\Xi_{cc}^{++}$ comes
from the experiment \cite{Aaij:2017ueg} and other masses of doubly
heavy baryons are predictions of the Lattice QCD \cite{Brown:2014ena}.
Masses for baryons with a single heavy quark are taken from Particle
Data Group~\cite{Olive:2016xmw,Tanabashi:2018oca}. Masses of the
negative parity baryons presented in Eq. \eqref{eq:Fi_plus_plus}
are collected in Table \ref{Tab:mass_if_negative} \cite{Wang:2010it,Roberts:2007ni}.

When arriving at the predictions of the branching ratios, the lifetimes
of the initial doubly-heavy baryons are aslo needed. They are collected
in Table~\ref{Tab:lifetime}, in which the lifetime of $\Xi_{cc}^{++}$
comes from the experiment \cite{Aaij:2018wzf}, and other results
are the theoretical predictions \cite{Karliner:2014gca,Kiselev:2001fw,Cheng:2018mwu}.

\begin{table}
\caption{``Decay constants\char`\"{} (pole residues) for the doubly-heavy
and singly-heavy hadrons as well as their masses. Results for charmed
and bottom baryons are taken from Refs.~\cite{Wang:2010fq,Wang:2009cr},
while for doubly-heavy baryons, we have updated the pole residues
in Ref.~\cite{Hu:2017dzi} using the same inputs as those in this
work. The factor $\sqrt{2}$ arises from the convention differences
in the definitions of the interpolating current for baryon. The mass
of $\Xi_{cc}^{++}$ comes from the experiment \cite{Aaij:2017ueg}
and other masses of doubly heavy baryons are predictions of the Lattice
QCD \cite{Brown:2014ena}. Masses for baryons with a single heavy
quark are taken from Particle Data Group~\cite{Olive:2016xmw,Tanabashi:2018oca}.}
\label{Tab:decay_constant} %
\begin{tabular}{c|ccc|c}
\hline 
 & $T^{2}({\rm GeV}^{2})$  & $\sqrt{s_{0}}({\rm GeV})$  & $M({\rm GeV})$  & $\lambda({\rm GeV}^{3})$ \tabularnewline
\hline 
$\Lambda_{c}$  & $1.7-2.7$  & $3.1\pm0.1$  & $2.286$  & $\sqrt{2}(0.022\pm0.003)$ \tabularnewline
$\Xi_{c}$  & $1.9-2.9$  & $3.2\pm0.1$  & $2.468$  & $\sqrt{2}(0.027\pm0.004)$ \tabularnewline
$\Lambda_{b}$  & $4.3-5.3$  & $6.5\pm0.1$  & $5.620$  & $\sqrt{2}(0.028\pm0.004)$ \tabularnewline
$\Xi_{b}$  & $4.4-5.4$  & $6.5\pm0.1$  & $5.793$  & $\sqrt{2}(0.034\pm0.006)$ \tabularnewline
\hline 
$\Sigma_{c}$  & $1.8-2.8$  & $3.2\pm0.1$  & $2.454$  & $\sqrt{2}(0.046\pm0.006)$ \tabularnewline
$\Xi_{c}^{\prime}$  & $2.0-3.0$  & $3.3\pm0.1$  & $2.576$  & $\sqrt{2}(0.054\pm0.007)$ \tabularnewline
$\Omega_{c}$  & $2.2-3.2$  & $3.4\pm0.1$  & $2.695$  & $0.089\pm0.013$ \tabularnewline
$\Sigma_{b}$  & $4.6-5.6$  & $6.6\pm0.1$  & $5.814$  & $\sqrt{2}(0.062\pm0.010)$ \tabularnewline
$\Xi_{b}^{\prime}$  & $4.9-5.9$  & $6.7\pm0.1$  & $5.935$  & $\sqrt{2}(0.074\pm0.011)$ \tabularnewline
$\Omega_{b}$  & $5.2-6.2$  & $6.8\pm0.1$  & $6.046$  & $0.123\pm0.020$ \tabularnewline
\hline 
\hline 
$\Xi_{cc}$  & $2.4-3.4$  & $4.1\pm0.1$  & $3.621$  & $0.109\pm0.020$ \tabularnewline
$\Omega_{cc}$  & $2.6-3.6$  & $4.3\pm0.1$  & $3.738\pm0.028$  & $0.129\pm0.024$ \tabularnewline
$\Xi_{bb}$  & $6.8-7.8$  & $10.6\pm0.1$  & $10.143\pm0.038$  & $0.199\pm0.052$ \tabularnewline
$\Omega_{bb}$  & $7.2-8.2$  & $10.8\pm0.1$  & $10.273\pm0.034$  & $0.253\pm0.062$ \tabularnewline
$\Xi_{bc}$  & $4.2-5.2$  & $7.4\pm0.1$  & $6.943\pm0.043$  & $0.150\pm0.035$ \tabularnewline
$\Omega_{bc}$  & $4.5-5.5$  & $7.6\pm0.1$  & $6.998\pm0.034$  & $0.168\pm0.038$ \tabularnewline
\hline 
\end{tabular}
\end{table}
\begin{table}[!htb]
\caption{Masses (in units of GeV) of the negative parity baryons \cite{Wang:2010it,Roberts:2007ni}.}
\label{Tab:mass_if_negative}

\begin{tabular}{c|c|c|c|c|c|c}
\hline 
Baryon  & $\Xi_{cc}(\frac{1}{2}^{-})$  & $\Omega_{cc}(\frac{1}{2}^{-})$  & $\Xi_{bc}(\frac{1}{2}^{-})$  & $\Omega_{bc}(\frac{1}{2}^{-})$  & $\Xi_{bb}(\frac{1}{2}^{-})$  & $\Omega_{bb}(\frac{1}{2}^{-})$ \tabularnewline
\hline 
Mass  & $3.77$ \cite{Wang:2010it}  & $3.91$ \cite{Wang:2010it}  & $7.231$ \cite{Roberts:2007ni}  & $7.346$ \cite{Roberts:2007ni}  & $10.38$ \cite{Wang:2010it}  & $10.53$ \cite{Wang:2010it} \tabularnewline
\hline 
\end{tabular}%
\\
\begin{tabular}{c|c|c|c|c|c}
\hline 
Baryon  & $\Lambda_{c}(\frac{1}{2}^{-})$  & $\Xi_{c}(\frac{1}{2}^{-})$  & $\Sigma_{c}(\frac{1}{2}^{-})$  & $\Xi_{c}^{\prime}(\frac{1}{2}^{-})$  & $\Omega_{c}(\frac{1}{2}^{-})$\tabularnewline
\hline 
Mass  & $2.592$ \cite{Roberts:2007ni}  & $2.789$ \cite{Roberts:2007ni}  & $2.74$ \cite{Wang:2010it}  & $2.87$ \cite{Wang:2010it}  & $2.98$ \cite{Wang:2010it} \tabularnewline
\hline 
Baryon  & $\Lambda_{b}(\frac{1}{2}^{-})$  & $\Xi_{b}(\frac{1}{2}^{-})$  & $\Sigma_{b}(\frac{1}{2}^{-})$  & $\Xi_{b}^{\prime}(\frac{1}{2}^{-})$  & $\Omega_{b}(\frac{1}{2}^{-})$\tabularnewline
\hline 
Mass  & $5.912$ \cite{Roberts:2007ni}  & $6.108$ \cite{Roberts:2007ni}  & $6.00$ \cite{Wang:2010it}  & $6.14$ \cite{Wang:2010it}  & $6.27$ \cite{Wang:2010it} \tabularnewline
\hline 
\end{tabular}
\end{table}
\begin{table}[!htb]
\caption{Lifetimes (in units of fs) of doubly-heavy baryons. The lifetime of
$\Xi_{cc}^{++}$ comes from the experiment \cite{Aaij:2018wzf}, and
other results are theoretical predictions \cite{Karliner:2014gca,Kiselev:2001fw,Cheng:2018mwu}.}
\label{Tab:lifetime} %
\begin{tabular}{c|c|c|c|c|c|c|c|c|c}
\hline 
Baryon  & $\Xi_{cc}^{++}$  & $\Xi_{cc}^{+}$  & $\Omega_{cc}^{+}$  & $\Xi_{bc}^{+}$  & $\Xi_{bc}^{0}$  & $\Omega_{bc}^{0}$  & $\Xi_{bb}^{0}$  & $\Xi_{bb}^{-}$  & $\Omega_{bb}^{-}$ \tabularnewline
\hline 
Lifetime  & $256$ \cite{Aaij:2018wzf}  & $44$ \cite{Cheng:2018mwu}  & $206$ \cite{Cheng:2018mwu}  & $244$ \cite{Karliner:2014gca}  & $93$ \cite{Karliner:2014gca}  & $220$ \cite{Kiselev:2001fw}  & $370$ \cite{Karliner:2014gca}  & $370$ \cite{Karliner:2014gca}  & $800$ \cite{Kiselev:2001fw}\tabularnewline
\hline 
\end{tabular}
\end{table}

The threshold parameters $\sqrt{s_{1,2}^{0}}$ are taken from Table
\ref{Tab:decay_constant}, which are essentially about $0.5$ GeV
higher than the corresponding baryon mass \cite{Wang:2012kw}. We employ the following equation from Ref.~\cite{Ball:1991bs}
to simplify the selection of Borel mass parameters:
\begin{equation}
\frac{T_{1}^{2}}{T_{2}^{2}}\approx\frac{M_{1}^{2}-m_{1}^{2}}{M_{2}^{2}-m_{1}^{\prime2}},\label{eq:T12_T22_relation}
\end{equation}
where $M_{1(2)}$ is the mass of the initial (final) baryon and $m_{1}^{(\prime)}$
is the mass of the initial (final) quark. To determine the window
of the Borel parameter $T_{1}^{2}$, the criteria of pole dominance
\begin{equation}
r\equiv\frac{\int^{s_{1}^{0}}ds_{1}\int^{s_{2}^{0}}ds_{2}\rho^{{\rm QCD}}(s_{1},s_{2},q^{2})\exp\left(-s_{1}/T_{1}^{2}-s_{2}/T_{2}^{2}\right)}{\int^{\infty}ds_{1}\int^{\infty}ds_{2}\rho^{{\rm QCD}}(s_{1},s_{2},q^{2})\exp\left(-s_{1}/T_{1}^{2}-s_{2}/T_{2}^{2}\right)}\gtrsim0.5\label{eq:pole_over_all}
\end{equation}
and OPE convergence are invoked. For the latter, the reader can refer to Table \ref{Tab:contribution_dim_0354}. The obtained windows for $T_{1}^{2}$
can be seen in Table \ref{Tab:T12_window}. In Table \ref{Tab:error_estimate}, we have evaluated all the error
sources for the form factors of $\Xi_{cc}^{++}\to\Sigma_{c}^{+}$. One can
see that the Borel parameter dependence is weak.

\begin{figure}
\includegraphics[width=\columnwidth]{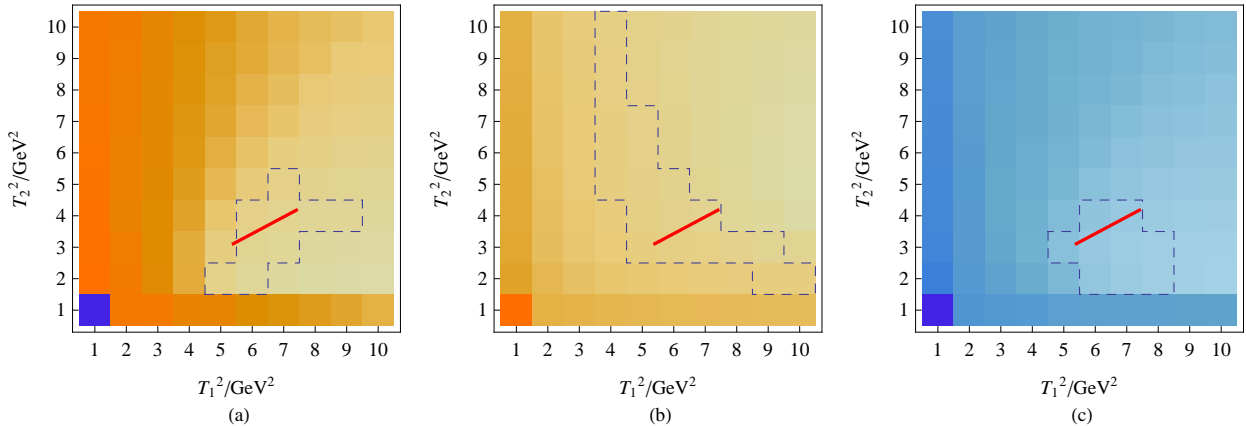}	
\caption{$F_{1, 2, 3}$ at $q^{2}=0$ as functions of the Borel parameters $T_{1}^{2}$ and $T_{2}^{2}$ in the process of $\Xi_{cc}^{++}\to\Sigma_{c}^{+}$, where $T_{1}^{2}$ and $T_{2}^{2}$ are taken as free parameters. Positive and negative values are respectively displayed as reddish ($F_{1,2}$) and bluish ($F_{3}$), and the greater the absolute value for the form factors $F_{i}(0)$, the darker the color. The allowed Borel regions are enclosed by the dashed contours. To determine these regions, three criteria have been applied, as can be seen in the text. The red line segment determined by Eq. (\ref{eq:T12_T22_relation}), which is adopted in this work, is also shown on each figure. }
\label{fig:Borel} 
\end{figure}

More comments on the selection of the Borel parameters are in order. $T_{1}^{2}$ and $T_{2}^{2}$ are in fact free parameters in the dispersion integral. To investigate the dependence on the Borel parameters, we take the $\Xi_{cc}^{++}\to\Sigma_{c}^{+}$ process as an example. First, we calculate the form factors $F_{1, 2, 3}(0)$ as functions of $T_{1}^{2}$ and $T_{2}^{2}$ in the square region $[1,10]$ GeV$^{2}\times[1,10]$ GeV$^{2}$. Then the obtained results are represented graphically in Fig. \ref{fig:Borel}, where the positive and negative values are respectively displayed as reddish and bluish, and the greater the absolute value for the form factors $F_{1, 2, 3}(0)$, the darker the color. In the end, the following three criteria are employed to determine the Borel region:
\begin{itemize}
\item The pole dominance. See Eq. (\ref{eq:pole_over_all}).
\item OPE convergence. This can be achieved by demanding that the contribution from the quark-gluon condensate (dim-5) is less than, for example, 10\%.
\item Stability of the quantity within the Borel region. This can be read directly from Fig. \ref{fig:Borel}. 
\end{itemize}
More details can be found in Table \ref{Tab:criteria_and_comparison}. In Fig. \ref{fig:Borel}, we also show the line segment determined by Eq. (\ref{eq:T12_T22_relation}). It can be seen that, the simplified Eq. (\ref{eq:T12_T22_relation}) is still a good approximation, and a quantitative comparison between these two different ways to determine Borel parameters can be seen in Table \ref{Tab:criteria_and_comparison}.

\begin{table}
\caption{
The quantitative criteria of the pole dominance and OPE convergence, and a comparison of the results of $F_{i}(0)$ obtained by these two different ways to determine Borel parameters. When the Borel parameters are taken as free, we average the values of $F_{i}(0)$ within the Borel region in Fig. \ref{fig:Borel}, and when Eq. (\ref{eq:T12_T22_relation}) is used, the value evaluated at the midpoint of the line segment in Fig. \ref{fig:Borel} is shown. The process of $\Xi_{cc}^{++}\to\Sigma_{c}^{+}$ is considered.
}
\label{Tab:criteria_and_comparison} %
\begin{tabular}{c|ccc}
\hline 
 & $F_{1}(0)$  & $F_{2}(0)$  & $F_{3}(0)$ \tabularnewline
\hline 
The pole dominance $r$  & $>0.5$  & $>0.5$  & $>0.45$ \tabularnewline
OPE convergence dim-5/total & $<5\%$  & $<10\%$  & $<8\%$ \tabularnewline
\hline 
Free Borel parameters & $1.126$ & $0.638$ & $-2.008$\tabularnewline
Constrained  Borel parameters & $1.147$ & $0.641$ & $-2.059$\tabularnewline
\hline 
\end{tabular}
\end{table}

\begin{table}
\caption{Contributions to form factors from dim-0, 3, 5 and the gluon-gluon condensate shown in Fig.~\ref{fig:GG2} for the $\Xi_{cc}^{++}\to\Sigma_{c}^{+}$
transition with $T_{1}^{2}$ taking as the central value $5.9\ {\rm GeV}^{2}$.}
\label{Tab:contribution_dim_0354}
\begin{tabular}{c|cccccc}
\hline 
 & $F_{1}(0)$ & $F_{2}(0)$ & $F_{3}(0)$ & $G_{1}(0)$ & $G_{2}(0)$ & $G_{3}(0)$\tabularnewline
\hline 
dim-0 & $0.507$ & $0.194$ & $-0.824$ & $-1.162$ & $0.627$ & $0.205$\tabularnewline
dim-3 & $0.606$ & $0.391$ & $-1.129$ & $-1.605$ & $1.099$ & $0.211$\tabularnewline
dim-5 & $0.034$ & $0.056$ & $-0.106$ & $-0.291$ & $0.217$ & $0.012$\tabularnewline
\hline 
Fig.~\ref{fig:GG2}  & $-0.007$ & $-0.008$ & $0.012$ & - - & - - & - -\tabularnewline
\hline 
\end{tabular}
\end{table}
\begin{table}
\caption{The windows of the Borel parameter $T_{1}^{2}$ and the range of $r$
in Eq. (\ref{eq:pole_over_all}) for the form factors in different transitions.
$T_{2}^{2}$ is determined by Eq. (\ref{eq:T12_T22_relation}). The
momentum transfer squared $q^{2}$ is taken at $-0.5\ {\rm GeV}^{2}$ ($-5\ {\rm GeV}^{2}$)
for the case of $c$ ($b$) quark decay. The central value of $T_{1}^{2}$
will be taken as the midpoint of the interval.}
\label{Tab:T12_window} %
\begin{tabular}{c|cccc}
\hline 
Transition & $T_{1}^{2}({\rm GeV}^{2})$  & $F_{1}$ & $F_{2}$ & $F_{3}$\tabularnewline
\hline 
$\Xi_{cc}\to\Sigma_{c}$  & $4.9-6.9$  & {\footnotesize{}$(61-82)\%$ } & {\footnotesize{}$(56-80)\%$ } & {\footnotesize{}$(51-76)\%$ }\tabularnewline
$\Xi_{bc}\to\Sigma_{b}$ & $10.2-12.2$  & {\footnotesize{}$(70-93)\%$ } & {\footnotesize{}$(50-68)\%$ } & {\footnotesize{}$(54-75)\%$ }\tabularnewline
$\Xi_{bc}\to\Sigma_{c}$ & $9.8-11.8$  & {\footnotesize{}$(53-70)\%$ } & {\footnotesize{}$(57-67)\%$ } & {\footnotesize{}$(50-61)\%$ }\tabularnewline
$\Xi_{bb}\to\Sigma_{b}$ & $11.9-13.9$  & {\footnotesize{}$(51-58)\%$ } & {\footnotesize{}$(54-61)\%$ } & {\footnotesize{}$(50-57)\%$ }\tabularnewline
\hline 
$\Xi_{cc}\to\Lambda_{c}$  & $5.7-7.7$  & {\footnotesize{}$(86-90)\%$ } & {\footnotesize{}$(72-85)\%$ } & {\footnotesize{}$(51-73)\%$ }\tabularnewline
$\Xi_{bc}\to\Lambda_{b}$ & $11.6-13.6$  & {\footnotesize{}$(72-89)\%$ } & {\footnotesize{}$(69-91)\%$ } & {\footnotesize{}$(51-68)\%$ }\tabularnewline
$\Xi_{bc}\to\Lambda_{c}$ & $10.4-12.4$  & {\footnotesize{}$(65-70)\%$ } & {\footnotesize{}$(58-66)\%$ } & {\footnotesize{}$(50-60)\%$ }\tabularnewline
$\Xi_{bb}\to\Lambda_{b}$ & $10.9-12.9$  & {\footnotesize{}$(51-59)\%$ } & {\footnotesize{}$(52-59)\%$ } & {\footnotesize{}$(50-57)\%$ }\tabularnewline
\hline 
\end{tabular}
\end{table}
\begin{table}
\caption{The form factors for the $cc$ sector. Eq. (\ref{eq:fit_formula_1})
is adopted as the fit formula. The results for $\Xi_{cc}\to\Sigma_{c}$
correspond to $\Xi_{cc}^{++}\to\Sigma_{c}^{+}$. A factor $\sqrt{2}$
should be multiplied to $F(0)$ for $\Xi_{cc}^{+}\to\Sigma_{c}^{0}$.
The form factor $f_{2}$ in the $\Xi_{cc}\to\Xi_{c}$ and $\Omega_{cc}\to\Xi_{c}$
transitions can not be fitted well, the corresponding $(m_{{\rm fit}},\delta)$
are taken from those in the $\Xi_{cc}\to\Lambda_{c}$ transition. For $F(0)$, we have only considered the uncertainty from the heavy quark masses.}
\label{Tab:ff_cc}%
\begin{tabular}{c|c|c|c|c|c|c|c}
\hline 
$F$  & $F(0)$  & $m_{{\rm fit}}$  & $\delta$  & $F$  & $F(0)$  & $m_{{\rm fit}}$  & $\delta$ \tabularnewline
\hline 
$f_{1}^{\Xi_{cc}\to\Lambda_{c}}$  & $-0.63\pm0.20$  & $1.57$  & $0.08$  & $g_{1}^{\Xi_{cc}\to\Lambda_{c}}$  & $0.24\pm0.08$  & $2.27$  & $0.39$ \tabularnewline
$f_{2}^{\Xi_{cc}\to\Lambda_{c}}$  & $0.05\pm0.02$  & $2.43$  & $2.10$  & $g_{2}^{\Xi_{cc}\to\Lambda_{c}}$  & $-0.11\pm0.03$  & $1.54$  & $0.12$ \tabularnewline
$f_{3}^{\Xi_{cc}\to\Lambda_{c}}$  & $0.81\pm0.26$  & $1.34$  & $0.20$  & $g_{3}^{\Xi_{cc}\to\Lambda_{c}}$  & $-0.84\pm0.30$  & $1.34$  & $0.20$ \tabularnewline
\hline 
$f_{1}^{\Xi_{cc}\to\Xi_{c}}$  & $-0.69\pm0.23$  & $1.54$  & $-0.01$  & $g_{1}^{\Xi_{cc}\to\Xi_{c}}$  & $0.25\pm0.08$  & $2.30$  & $0.39$ \tabularnewline
$f_{2}^{\Xi_{cc}\to\Xi_{c}}$  & $0.06\pm0.02$  & $2.43$  & $2.10$  & $g_{2}^{\Xi_{cc}\to\Xi_{c}}$  & $-0.14\pm0.04$  & $1.54$  & $0.21$ \tabularnewline
$f_{3}^{\Xi_{cc}\to\Xi_{c}}$  & $0.91\pm0.30$  & $1.30$  & $0.12$  & $g_{3}^{\Xi_{cc}\to\Xi_{c}}$  & $-0.92\pm0.31$  & $1.34$  & $0.22$ \tabularnewline
\hline 
$f_{1}^{\Omega_{cc}\to\Xi_{c}}$  & $-0.67\pm0.21$  & $1.66$  & $0.21$  & $g_{1}^{\Omega_{cc}\to\Xi_{c}}$  & $0.25\pm0.08$  & $2.34$  & $0.38$ \tabularnewline
$f_{2}^{\Omega_{cc}\to\Xi_{c}}$  & $0.06\pm0.02$  & $2.43$  & $2.10$  & $g_{2}^{\Omega_{cc}\to\Xi_{c}}$  & $-0.12\pm0.03$  & $1.51$  & $-0.05$ \tabularnewline
$f_{3}^{\Omega_{cc}\to\Xi_{c}}$  & $0.84\pm0.26$  & $1.37$  & $0.19$  & $g_{3}^{\Omega_{cc}\to\Xi_{c}}$  & $-0.89\pm0.30$  & $1.35$  & $0.12$ \tabularnewline
\hline 
$f_{1}^{\Xi_{cc}\to\Sigma_{c}}$  & $-0.30\pm0.07$  & $1.76$  & $-0.65$  & $g_{1}^{\Xi_{cc}\to\Sigma_{c}}$  & $0.46\pm0.15$  & $2.29$  & $0.41$ \tabularnewline
$f_{2}^{\Xi_{cc}\to\Sigma_{c}}$  & $1.05\pm0.38$  & $1.57$  & $0.23$  & $g_{2}^{\Xi_{cc}\to\Sigma_{c}}$  & $-0.09\pm0.01$  & $1.20$  & $1.59$ \tabularnewline
$f_{3}^{\Xi_{cc}\to\Sigma_{c}}$  & $0.10\pm0.00$  & $1.00$  & $0.78$  & $g_{3}^{\Xi_{cc}\to\Sigma_{c}}$  & $-2.96\pm1.13$  & $1.34$  & $0.16$ \tabularnewline
\hline 
$f_{1}^{\Xi_{cc}\to\Xi_{c}^{\prime}}$  & $-0.31\pm0.06$  & $2.25$  & $1.08$  & $g_{1}^{\Xi_{cc}\to\Xi_{c}^{\prime}}$  & $0.50\pm0.17$  & $2.28$  & $0.42$ \tabularnewline
$f_{2}^{\Xi_{cc}\to\Xi_{c}^{\prime}}$  & $1.10\pm0.40$  & $1.54$  & $0.12$  & $g_{2}^{\Xi_{cc}\to\Xi_{c}^{\prime}}$  & $-0.17\pm0.03$  & $1.14$  & $0.48$ \tabularnewline
$f_{3}^{\Xi_{cc}\to\Xi_{c}^{\prime}}$  & $0.15\pm0.02$  & $1.02$  & $0.44$  & $g_{3}^{\Xi_{cc}\to\Xi_{c}^{\prime}}$  & $-3.09\pm1.18$  & $1.34$  & $0.15$ \tabularnewline
\hline 
$f_{1}^{\Omega_{cc}\to\Xi_{c}^{\prime}}$  & $-0.28\pm0.05$  & $2.07$  & $-0.60$  & $g_{1}^{\Omega_{cc}\to\Xi_{c}^{\prime}}$  & $0.49\pm0.16$  & $2.20$  & $-0.07$ \tabularnewline
$f_{2}^{\Omega_{cc}\to\Xi_{c}^{\prime}}$  & $1.13\pm0.40$  & $1.59$  & $0.19$  & $g_{2}^{\Omega_{cc}\to\Xi_{c}^{\prime}}$  & $-0.08\pm0.01$  & $1.22$  & $2.60$ \tabularnewline
$f_{3}^{\Omega_{cc}\to\Xi_{c}^{\prime}}$  & $0.07\pm0.00$  & $1.15$  & $4.13$  & $g_{3}^{\Omega_{cc}\to\Xi_{c}^{\prime}}$  & $-3.20\pm1.19$  & $1.37$  & $0.16$ \tabularnewline
\hline 
$f_{1}^{\Omega_{cc}\to\Omega_{c}}$  & $-0.42\pm0.08$  & $1.78$  & $-0.96$  & $g_{1}^{\Omega_{cc}\to\Omega_{c}}$  & $0.74\pm0.25$  & $2.37$  & $0.54$ \tabularnewline
$f_{2}^{\Omega_{cc}\to\Omega_{c}}$  & $1.66\pm0.58$  & $1.65$  & $0.36$  & $g_{2}^{\Omega_{cc}\to\Omega_{c}}$  & $-0.19\pm0.03$  & $1.59$  & $3.77$ \tabularnewline
$f_{3}^{\Omega_{cc}\to\Omega_{c}}$  & $0.16\pm0.01$  & $1.20$  & $1.81$  & $g_{3}^{\Omega_{cc}\to\Omega_{c}}$  & $-4.72\pm1.76$  & $1.36$  & $0.15$ \tabularnewline
\hline 
\end{tabular}
\end{table}

\begin{table}
\caption{The form factors for the $bb$ sector. Eq. (\ref{eq:fit_formula_1})
is adopted as the fit formula. The results for $\Xi_{bb}\to\Sigma_{b}$
correspond to $\Xi_{bb}^{-}\to\Sigma_{b}^{0}$. A factor $\sqrt{2}$
should be multiplied to $F(0)$ for $\Xi_{bb}^{0}\to\Sigma_{b}^{+}$. For $F(0)$, we have only considered the uncertainty from the heavy quark masses.}
\label{Tab:ff_bb}%
\begin{tabular}{c|c|c|c|c|c|c|c}
\hline 
$F$  & $F(0)$  & $m_{{\rm fit}}$  & $\delta$  & $F$  & $F(0)$  & $m_{{\rm fit}}$  & $\delta$ \tabularnewline
\hline 
$f_{1}^{\Xi_{bb}\to\Lambda_{b}}$  & $-0.072\pm0.041$  & $2.52$  & $0.39$  & $g_{1}^{\Xi_{bb}\to\Lambda_{b}}$  & $0.027\pm0.015$  & $2.65$  & $0.41$ \tabularnewline
$f_{2}^{\Xi_{bb}\to\Lambda_{b}}$  & $0.004\pm0.003$  & $2.62$  & $0.40$  & $g_{2}^{\Xi_{bb}\to\Lambda_{b}}$  & $-0.013\pm0.007$  & $2.47$  & $0.39$ \tabularnewline
$f_{3}^{\Xi_{bb}\to\Lambda_{b}}$  & $0.085\pm0.048$  & $2.41$  & $0.37$  & $g_{3}^{\Xi_{bb}\to\Lambda_{b}}$  & $-0.069\pm0.040$  & $2.42$  & $0.37$ \tabularnewline
\hline 
$f_{1}^{\Omega_{bb}\to\Xi_{b}}$  & $-0.095\pm0.053$  & $2.66$  & $0.35$  & $g_{1}^{\Omega_{bb}\to\Xi_{b}}$  & $0.036\pm0.021$  & $2.81$  & $0.36$ \tabularnewline
$f_{2}^{\Omega_{bb}\to\Xi_{b}}$  & $0.006\pm0.004$  & $2.72$  & $0.37$  & $g_{2}^{\Omega_{bb}\to\Xi_{b}}$  & $-0.017\pm0.009$  & $2.60$  & $0.35$ \tabularnewline
$f_{3}^{\Omega_{bb}\to\Xi_{b}}$  & $0.112\pm0.063$  & $2.52$  & $0.35$  & $g_{3}^{\Omega_{bb}\to\Xi_{b}}$  & $-0.093\pm0.053$  & $2.53$  & $0.35$ \tabularnewline
\hline 
$f_{1}^{\Xi_{bb}\to\Sigma_{b}}$  & $-0.050\pm0.026$  & $2.89$  & $0.38$  & $g_{1}^{\Xi_{bb}\to\Sigma_{b}}$  & $0.060\pm0.032$  & $2.96$  & $0.39$ \tabularnewline
$f_{2}^{\Xi_{bb}\to\Sigma_{b}}$  & $0.149\pm0.082$  & $2.65$  & $0.37$  & $g_{2}^{\Xi_{bb}\to\Sigma_{b}}$  & $0.016\pm0.008$  & $3.24$  & $0.75$ \tabularnewline
$f_{3}^{\Xi_{bb}\to\Sigma_{b}}$  & $0.012\pm0.005$  & $2.35$  & $0.34$  & $g_{3}^{\Xi_{bb}\to\Sigma_{b}}$  & $-0.377\pm0.205$  & $2.60$  & $0.36$ \tabularnewline
\hline 
$f_{1}^{\Omega_{bb}\to\Xi_{b}^{\prime}}$  & $-0.057\pm0.028$  & $2.97$  & $0.39$  & $g_{1}^{\Omega_{bb}\to\Xi_{b}^{\prime}}$  & $0.072\pm0.037$  & $2.99$  & $0.41$ \tabularnewline
$f_{2}^{\Omega_{bb}\to\Xi_{b}^{\prime}}$  & $0.180\pm0.095$  & $2.70$  & $0.37$  & $g_{2}^{\Omega_{bb}\to\Xi_{b}^{\prime}}$  & $0.019\pm0.010$  & $3.69$  & $0.89$ \tabularnewline
$f_{3}^{\Omega_{bb}\to\Xi_{b}^{\prime}}$  & $0.012\pm0.005$  & $2.27$  & $0.36$  & $g_{3}^{\Omega_{bb}\to\Xi_{b}^{\prime}}$  & $-0.453\pm0.234$  & $2.65$  & $0.37$\tabularnewline
\hline 
\end{tabular}
\end{table}

\begin{table}
\caption{The form factors for the $bc$ sector with $c$ quark decay. Eq. (\ref{eq:fit_formula_1})
is adopted as the fit formula. The results for $\Xi_{bc}\to\Sigma_{b}$
correspond to $\Xi_{bc}^{+}\to\Sigma_{b}^{0}$. A factor $\sqrt{2}$
should be multiplied to $F(0)$ for $\Xi_{bc}^{0}\to\Sigma_{b}^{-}$.
The form factors $f_{1}$, $g_{1}$ and $g_{2}$ in the $\Xi_{bc}\to\Xi_{b}^{\prime}$
transition can not be fitted well, the corresponding $(m_{{\rm fit}},\delta)$
are taken from those in the $\Xi_{bc}\to\Sigma_{b}$ transition. Also,
the form factor $f_{1}$ in the $\Omega_{bc}\to\Omega_{b}$ transition
can not be fitted well, the corresponding $(m_{{\rm fit}},\delta)$
are taken from those in the $\Omega_{bc}\to\Xi_{b}^{\prime}$ transition. For $F(0)$, we have only considered the uncertainty from the heavy quark masses.}
\label{Tab:ff_bc_c}%
\begin{tabular}{c|c|c|c|c|c|c|c}
\hline 
$F$  & $F(0)$  & $m_{{\rm fit}}$  & $\delta$  & $F$  & $F(0)$  & $m_{{\rm fit}}$  & $\delta$ \tabularnewline
\hline 
$f_{1}^{\Xi_{bc}\to\Lambda_{b}}$  & $-0.45\pm0.15$  & $1.33$  & $0.06$  & $g_{1}^{\Xi_{bc}\to\Lambda_{b}}$  & $0.16\pm0.05$  & $1.88$  & $0.49$ \tabularnewline
$f_{2}^{\Xi_{bc}\to\Lambda_{b}}$  & $0.31\pm0.09$  & $1.58$  & $0.47$  & $g_{2}^{\Xi_{bc}\to\Lambda_{b}}$  & $-0.14\pm0.04$  & $1.28$  & $0.61$ \tabularnewline
$f_{3}^{\Xi_{bc}\to\Lambda_{b}}$  & $1.21\pm0.37$  & $1.23$  & $0.27$  & $g_{3}^{\Xi_{bc}\to\Lambda_{b}}$  & $-2.74\pm0.83$  & $1.32$  & $0.23$ \tabularnewline
\hline 
$f_{1}^{\Xi_{bc}\to\Xi_{b}}$  & $-0.45\pm0.14$  & $1.42$  & $0.26$  & $g_{1}^{\Xi_{bc}\to\Xi_{b}}$  & $0.16\pm0.05$  & $1.90$  & $0.45$ \tabularnewline
$f_{2}^{\Xi_{bc}\to\Xi_{b}}$  & $0.32\pm0.10$  & $1.46$  & $0.11$  & $g_{2}^{\Xi_{bc}\to\Xi_{b}}$  & $-0.15\pm0.05$  & $1.18$  & $0.28$ \tabularnewline
$f_{3}^{\Xi_{bc}\to\Xi_{b}}$  & $1.23\pm0.38$  & $1.24$  & $0.28$  & $g_{3}^{\Xi_{bc}\to\Xi_{b}}$  & $-2.79\pm0.83$  & $1.31$  & $0.21$ \tabularnewline
\hline 
$f_{1}^{\Omega_{bc}\to\Xi_{b}}$  & $-0.44\pm0.13$  & $1.44$  & $0.21$  & $g_{1}^{\Omega_{bc}\to\Xi_{b}}$  & $0.16\pm0.05$  & $1.95$  & $0.48$ \tabularnewline
$f_{2}^{\Omega_{bc}\to\Xi_{b}}$  & $0.31\pm0.09$  & $1.54$  & $0.20$  & $g_{2}^{\Omega_{bc}\to\Xi_{b}}$  & $-0.12\pm0.04$  & $1.18$  & $0.32$ \tabularnewline
$f_{3}^{\Omega_{bc}\to\Xi_{b}}$  & $1.12\pm0.33$  & $1.28$  & $0.29$  & $g_{3}^{\Omega_{bc}\to\Xi_{b}}$  & $-2.62\pm0.76$  & $1.38$  & $0.25$ \tabularnewline
\hline 
$f_{1}^{\Xi_{bc}\to\Sigma_{b}}$  & $-0.23\pm0.06$  & $1.70$  & $0.67$  & $g_{1}^{\Xi_{bc}\to\Sigma_{b}}$  & $0.33\pm0.11$  & $1.73$  & $0.13$ \tabularnewline
$f_{2}^{\Xi_{bc}\to\Sigma_{b}}$  & $1.51\pm0.50$  & $1.39$  & $0.24$  & $g_{2}^{\Xi_{bc}\to\Sigma_{b}}$  & $-0.39\pm0.12$  & $1.09$  & $0.13$ \tabularnewline
$f_{3}^{\Xi_{bc}\to\Sigma_{b}}$  & $0.38\pm0.11$  & $1.04$  & $0.25$  & $g_{3}^{\Xi_{bc}\to\Sigma_{b}}$  & $-8.24\pm2.97$  & $1.24$  & $0.31$ \tabularnewline
\hline 
$f_{1}^{\Xi_{bc}\to\Xi_{b}^{\prime}}$  & $-0.24\pm0.06$  & $1.70$  & $0.67$  & $g_{1}^{\Xi_{bc}\to\Xi_{b}^{\prime}}$  & $0.35\pm0.11$  & $1.73$  & $0.13$ \tabularnewline
$f_{2}^{\Xi_{bc}\to\Xi_{b}^{\prime}}$  & $1.56\pm0.51$  & $1.48$  & $0.51$  & $g_{2}^{\Xi_{bc}\to\Xi_{b}^{\prime}}$  & $-0.46\pm0.12$  & $1.09$  & $0.13$ \tabularnewline
$f_{3}^{\Xi_{bc}\to\Xi_{b}^{\prime}}$  & $0.43\pm0.12$  & $1.09$  & $0.30$  & $g_{3}^{\Xi_{bc}\to\Xi_{b}^{\prime}}$  & $-8.44\pm3.09$  & $1.23$  & $0.23$ \tabularnewline
\hline 
$f_{1}^{\Omega_{bc}\to\Xi_{b}^{\prime}}$  & $-0.23\pm0.07$  & $1.66$  & $0.31$  & $g_{1}^{\Omega_{bc}\to\Xi_{b}^{\prime}}$  & $0.34\pm0.11$  & $1.89$  & $0.43$ \tabularnewline
$f_{2}^{\Omega_{bc}\to\Xi_{b}^{\prime}}$  & $1.56\pm0.50$  & $1.45$  & $0.30$  & $g_{2}^{\Omega_{bc}\to\Xi_{b}^{\prime}}$  & $-0.34\pm0.09$  & $1.23$  & $0.34$ \tabularnewline
$f_{3}^{\Omega_{bc}\to\Xi_{b}^{\prime}}$  & $0.35\pm0.09$  & $1.10$  & $0.31$  & $g_{3}^{\Omega_{bc}\to\Xi_{b}^{\prime}}$  & $-8.55\pm2.93$  & $1.28$  & $0.34$ \tabularnewline
\hline 
$f_{1}^{\Omega_{bc}\to\Omega_{b}}$  & $-0.32\pm0.07$  & $1.66$  & $0.31$  & $g_{1}^{\Omega_{bc}\to\Omega_{b}}$  & $0.51\pm0.17$  & $1.95$  & $0.59$ \tabularnewline
$f_{2}^{\Omega_{bc}\to\Omega_{b}}$  & $2.29\pm0.74$  & $1.46$  & $0.28$  & $g_{2}^{\Omega_{bc}\to\Omega_{b}}$  & $-0.60\pm0.18$  & $1.21$  & $0.28$ \tabularnewline
$f_{3}^{\Omega_{bc}\to\Omega_{b}}$  & $0.58\pm0.17$  & $1.08$  & $0.24$  & $g_{3}^{\Omega_{bc}\to\Omega_{b}}$  & $-12.50\pm4.38$  & $1.24$  & $0.21$ \tabularnewline
\hline 
\end{tabular}
\end{table}

\begin{table}
\caption{The form factors for the $bc$ sector with $b$ quark decay. Eq. (\ref{eq:fit_formula_1}) is adopted as the fit formula.
The results for $\Xi_{bc}\to\Sigma_{c}$ correspond to $\Xi_{bc}^{0}\to\Sigma_{c}^{+}$.
A factor $\sqrt{2}$ should be multiplied to $F(0)$ for $\Xi_{bc}^{+}\to\Sigma_{c}^{++}$.
The form factor $g_{1}$ in the $\Xi_{bc}\to\Lambda_{c}$ transition
can not be fitted well, the corresponding $(m_{{\rm fit}},\delta)$
are taken from those in the $\Omega_{bc}\to\Xi_{c}$ transition. For $F(0)$, we have only considered the uncertainty from the heavy quark masses.}
\label{Tab:ff_bc_b}%
\begin{tabular}{c|c|c|c|c|c|c|c}
\hline 
$F$  & $F(0)$  & $m_{{\rm fit}}$  & $\delta$  & $F$  & $F(0)$  & $m_{{\rm fit}}$  & $\delta$ \tabularnewline
\hline 
$f_{1}^{\Xi_{bc}\to\Lambda_{c}}$  & $-0.141\pm0.052$  & $3.56$  & $0.28$  & $g_{1}^{\Xi_{bc}\to\Lambda_{c}}$  & $0.067\pm0.024$  & $4.06$  & $0.37$ \tabularnewline
$f_{2}^{\Xi_{bc}\to\Lambda_{c}}$  & $-0.040\pm0.015$  & $3.42$  & $0.34$  & $g_{2}^{\Xi_{bc}\to\Lambda_{c}}$  & $-0.037\pm0.013$  & $3.62$  & $0.37$ \tabularnewline
$f_{3}^{\Xi_{bc}\to\Lambda_{c}}$  & $0.108\pm0.039$  & $3.29$  & $0.34$  & $g_{3}^{\Xi_{bc}\to\Lambda_{c}}$  & $-0.006\pm0.003$  & $2.25$  & $0.36$ \tabularnewline
\hline 
$f_{1}^{\Omega_{bc}\to\Xi_{c}}$  & $-0.172\pm0.059$  & $3.64$  & $0.33$  & $g_{1}^{\Omega_{bc}\to\Xi_{c}}$  & $0.079\pm0.027$  & $4.06$  & $0.37$ \tabularnewline
$f_{2}^{\Omega_{bc}\to\Xi_{c}}$  & $-0.047\pm0.017$  & $3.53$  & $0.34$  & $g_{2}^{\Omega_{bc}\to\Xi_{c}}$  & $-0.043\pm0.014$  & $3.80$  & $0.36$ \tabularnewline
$f_{3}^{\Omega_{bc}\to\Xi_{c}}$  & $0.130\pm0.044$  & $3.38$  & $0.34$  & $g_{3}^{\Omega_{bc}\to\Xi_{c}}$  & $-0.011\pm0.004$  & $2.42$  & $0.37$ \tabularnewline
\hline 
$f_{1}^{\Xi_{bc}\to\Sigma_{c}}$  & $-0.069\pm0.022$  & $4.84$  & $0.40$  & $g_{1}^{\Xi_{bc}\to\Sigma_{c}}$  & $0.088\pm0.032$  & $4.71$  & $0.38$ \tabularnewline
$f_{2}^{\Xi_{bc}\to\Sigma_{c}}$  & $0.159\pm0.058$  & $3.52$  & $0.33$  & $g_{2}^{\Xi_{bc}\to\Sigma_{c}}$  & $0.059\pm0.021$  & $3.79$  & $0.41$ \tabularnewline
$f_{3}^{\Xi_{bc}\to\Sigma_{c}}$  & $-0.036\pm0.015$  & $3.88$  & $0.44$  & $g_{3}^{\Xi_{bc}\to\Sigma_{c}}$  & $-0.257\pm0.089$  & $3.42$  & $0.34$ \tabularnewline
\hline 
$f_{1}^{\Omega_{bc}\to\Xi_{c}^{\prime}}$  & $-0.076\pm0.022$  & $5.08$  & $0.33$  & $g_{1}^{\Omega_{bc}\to\Xi_{c}^{\prime}}$  & $0.101\pm0.035$  & $4.77$  & $0.31$ \tabularnewline
$f_{2}^{\Omega_{bc}\to\Xi_{c}^{\prime}}$  & $0.179\pm0.062$  & $3.60$  & $0.34$  & $g_{2}^{\Omega_{bc}\to\Xi_{c}^{\prime}}$  & $0.063\pm0.021$  & $3.99$  & $0.43$ \tabularnewline
$f_{3}^{\Omega_{bc}\to\Xi_{c}^{\prime}}$  & $-0.040\pm0.016$  & $4.02$  & $0.46$  & $g_{3}^{\Omega_{bc}\to\Xi_{c}^{\prime}}$  & $-0.286\pm0.093$  & $3.51$  & $0.34$\tabularnewline
\hline 
\end{tabular}
\end{table}

Numerical results for the form factors are given in Tables \ref{Tab:ff_cc},
\ref{Tab:ff_bb}, \ref{Tab:ff_bc_c} and \ref{Tab:ff_bc_b} for the doubly-charmed, doubly-bottom and bottom-charm baryons.
In QCDSR, the OPE is applicable in the deep Euclidean region, where
$q^{2}\ll0$. In this work, we directly calculate the form factors
in the region $-1<q^{2}<0$ GeV$^{2}$ for the charm quark decay, and
$0<q^{2}<5$ GeV$^{2}$ for the bottom quark decay. In order to access
the $q^{2}$ distribution in the full kinematic region, the form factors
are extrapolated with a parametrization. We adopt the following
double-pole parameterization: 
\begin{equation}
F(q^{2})=\frac{F(0)}{1-\frac{q^{2}}{m_{{\rm fit}}^{2}}+\delta\left(\frac{q^{2}}{m_{{\rm fit}}^{2}}\right)^{2}}.\label{eq:fit_formula_1}
\end{equation}

A few  remarks are given in order. 
\begin{itemize}

\item We have also calculated part of the gluon-gluon condensate, shown in Fig.~\ref{fig:GG2}, for $\Xi_{cc}^{++}\to\Sigma_{c}^{+}$, and make a comparison with other contributions in Table \ref{Tab:contribution_dim_0354}.
From this table, it is plausible to conclude
the following pattern:
\begin{equation}
\text{dim-0}\sim\text{dim-3}\gg\text{dim-5}\gg\text{dim-4}.
\end{equation}
We intend to perform a more comprehensive analysis by including all the contributions from the gluon-gluon condensate in future.

\item The form factors $g_{i}$'s are determined in the following way. Rewrite
Eq. (\ref{eq:correlator_pole_formal}) as 
\begin{equation}
\Pi_{\mu}^{V,{\rm pole}}=\sum_{i=1}^{12}A_{i}^{V}e_{i\mu}^{V},\label{eq:correlator_V_pole_formal}
\end{equation}
and similarly write the pole contribution for the axial-vector current
correlation function as
\begin{equation}
\Pi_{\mu}^{A,{\rm pole}}=\sum_{i=1}^{12}A_{i}^{A}e_{i\mu}^{A},\label{eq:correlator_A_pole_formal}
\end{equation}
where $e_{i\mu}^{V}\equiv e_{i\mu}$ in Eq. (\ref{eq:e_i_mu}) and
\begin{eqnarray}
(e_{1,2,3,4}^{A})_{\mu} & \equiv & \{\slashed p_{2},1\}\times\{p_{1\mu}\gamma_{5}\}\times\{\slashed p_{1},1\},\nonumber \\
(e_{5,6,7,8}^{A})_{\mu} & \equiv & \{\slashed p_{2},1\}\times\{p_{2\mu}\gamma_{5}\}\times\{\slashed p_{1},1\},\nonumber \\
(e_{9,10,11,12}^{A})_{\mu} & \equiv & \{\slashed p_{2},1\}\times\{\gamma_{\mu}\gamma_{5}\}\times\{\slashed p_{1},1\}.
\end{eqnarray}
In the massless limit $m_{1}^{\prime}\to0$ and $m_{3}\to0$, one
can prove for the process of the final baryon belonging to the sextet:
\begin{eqnarray}
 &  & A_{i}^{A,\text{dim-0}}=-A_{i}^{V,\text{dim-0}},\quad A_{i}^{A,\text{dim-3}}=A_{i}^{V,{\rm \text{dim-3}}},\quad A_{i}^{A,\text{dim-5}}=A_{i}^{V,\text{dim-5}},\quad\text{ for }i\text{ odd},\nonumber \\
 &  & A_{i}^{A,{\rm \text{dim-0}}}=A_{i}^{V,{\rm \text{dim-0}}},\quad A_{i}^{A,{\rm \text{dim-3}}}=-A_{i}^{V,{\rm \text{dim-3}}},\quad A_{i}^{A,{\rm \text{dim-5}}}=-A_{i}^{V,{\rm \text{dim-5}}},\quad\text{ for }i\text{ even},
\end{eqnarray}
and for the process of the final baryon belonging to the anti-triplet:
\begin{eqnarray}
 &  & A_{i}^{A,\text{dim-0}}=A_{i}^{V,\text{dim-0}},\quad A_{i}^{A,\text{dim-3}}=-A_{i}^{V,{\rm \text{dim-3}}},\quad A_{i}^{A,\text{dim-5}}=-A_{i}^{V,\text{dim-5}},\quad\text{ for }i\text{ odd},\nonumber \\
 &  & A_{i}^{A,{\rm \text{dim-0}}}=-A_{i}^{V,{\rm \text{dim-0}}},\quad A_{i}^{A,{\rm \text{dim-3}}}=A_{i}^{V,{\rm \text{dim-3}}},\quad A_{i}^{A,{\rm \text{dim-5}}}=A_{i}^{V,{\rm \text{dim-5}}},\quad\text{ for }i\text{ even}.
\end{eqnarray}
Here $A_{i}^{A,\text{dim-0}}$ stands for the coefficient $A_{i}^{A}$
in Eq. (\ref{eq:correlator_A_pole_formal}) with the dim-0 correlation
function being considered only, and so forth.

\item The uncertainties of form factors arise from those from the heavy quark masses, Borel parameter $T_{1}^{2}$, thresholds $s_{1}^{0}$ and $s_{2}^{0}$, condensate parameters, pole residues and masses of initial and final baryons. A detail analysis can be found in Subsection \ref{subsec:Uncertainties} and Table \ref{Tab:error_estimate}. It can be seen from \ref{Tab:error_estimate} that, the uncertainty mainly comes from that of the heavy quark mass. Thus, in Tables \ref{Tab:ff_cc}, \ref{Tab:ff_bb}, \ref{Tab:ff_bc_c} and \ref{Tab:ff_bc_b}, we only list the uncertainties from the heavy quark masses.

\item In Table~\ref{Tab:ff_cc}, the $\Xi_{cc}\to\Sigma_{c}$ stands for
the $\Xi_{cc}^{++}\to\Sigma_{c}^{+}$ transition. A factor $\sqrt{2}$
should be added for the $\Xi_{cc}^{+}\to\Sigma_{c}^{0}$ transition.
This is consistent with the analysis based on the flavor SU(3) symmetry~\cite{Wang:2017azm}.
Similar arguments can also be found in Tables \ref{Tab:ff_bb}, \ref{Tab:ff_bc_c},
and \ref{Tab:ff_bc_b}.

\end{itemize}

\begin{table}
\caption{Comparison with the results of the light-front quark model (LFQM)~\cite{Wang:2017mqp},
the nonrelativistic quark model (NRQM) and the MIT bag model (MBM)~\cite{PerezMarcial:1989yh}
for the form factors of $\Xi_{cc}^{++}\to\Lambda_{c}^{+}$ and $\Xi_{cc}^{++}\to\Sigma_{c}^{+}$. The signs of the form factors of $\Xi_{cc}^{++}\to\Lambda_{c}^{+}$
in the LFQM have been flipped so that those of vector-current form
factors are the same as ours. For the same reason, all the results
from NRQM and MBM are multiplied by $-1$ except for $f_{2}$ and $g_{2}$,
whose sign conventions in Ref. \cite{PerezMarcial:1989yh} are different
from ours.}
\label{Tab:comparison_ff_cc} %
\begin{tabular}{c|c|c|c|c|c}
\hline 
Transition  & $F(0)$  & This work  & LFQM~\cite{Wang:2017mqp}  & NRQM ~\cite{PerezMarcial:1989yh}  & MBM ~\cite{PerezMarcial:1989yh} \tabularnewline
\hline 
$\Xi_{cc}^{++}\to\Lambda_{c}^{+}$  & $f_{1}(0)$  & $-0.63$  & $-0.79$  & $-0.36$  & $-0.45$\tabularnewline
 & $f_{2}(0)$  & $0.05$  & $0.01$  & $-0.14$  & $-0.01$\tabularnewline
 & $f_{3}(0)$  & $0.81$  & - -  & $-0.08$  & $0.28$\tabularnewline
 & $g_{1}(0)$  & $0.24$  & $-0.22$  & $-0.20$  & $-0.15$\tabularnewline
 & $g_{2}(0)$  & $-0.11$  & $0.05$  & $-0.01$  & $-0.01$\tabularnewline
 & $g_{3}(0)$  & $-0.84$  & - -  & $0.03$  & $0.70$\tabularnewline
\hline 
$\Xi_{cc}^{++}\to\Sigma_{c}^{+}$  & $f_{1}(0)$  & $-0.30$ & $-0.47$  & $-0.28$  & $-0.30$\tabularnewline
 & $f_{2}(0)$  & $1.05$ & $1.04$  & $0.14$  & $0.91$\tabularnewline
 & $f_{3}(0)$  & $0.10$ & - -  & $-0.10$  & $0.07$\tabularnewline
 & $g_{1}(0)$  & $0.46$ & $-0.62$  & $-0.70$  & $-0.56$\tabularnewline
 & $g_{2}(0)$  & $-0.09$ & $0.05$  & $-0.02$  & $0.05$\tabularnewline
 & $g_{3}(0)$  & $-2.96$ & - -  & $0.10$  & $2.59$\tabularnewline
\hline 
\end{tabular}
\end{table}
\begin{table}
\caption{Comparison with the results of the light-front quark model (LFQM)~\cite{Wang:2017mqp}
for the form factors of $\Xi_{bb}^{-}\to\Lambda_{b}^{0},\Sigma_{b}^{0}$,
$\Xi_{bc}^{+}\to\Lambda_{b}^{0},\Sigma_{b}^{0}$, and $\Xi_{bc}^{0}\to\Lambda_{c}^{+},\Sigma_{c}^{+}$.
The signs of the form factors of $\Xi_{bc}^{+}\to\Lambda_{b}^{0}$
in the LFQM have been flipped so that those of vector-current form
factors are the same as ours.}
\label{Tab:comparison_ff_bb_bc} \centering{}%
\begin{tabular}{c|c|c|c||c|c|c|c}
\hline 
Transition  & $F(0)$  & This work  & LFQM~\cite{Wang:2017mqp}  & Transition  & $F(0)$  & This work  & LFQM~\cite{Wang:2017mqp}\tabularnewline
\hline 
$\Xi_{bb}^{-}\to\Lambda_{b}^{0}$  & $f_{1}(0)$  & $-0.072$  & $-0.102$  & $\Xi_{bb}^{-}\to\Sigma_{b}^{0}$  & $f_{1}(0)$  & $-0.050$  & $-0.060$ \tabularnewline
 & $f_{2}(0)$  & $0.004$  & $0.001$  &  & $f_{2}(0)$  & $0.149$  & $0.150$ \tabularnewline
 & $f_{3}(0)$  & $0.085$  & - -  &  & $f_{3}(0)$  & $0.012$  & - -\tabularnewline
 & $g_{1}(0)$  & $0.027$  & $-0.036$  &  & $g_{1}(0)$  & $0.060$  & $-0.089$ \tabularnewline
 & $g_{2}(0)$  & $-0.013$  & $0.012$  &  & $g_{2}(0)$  & $0.016$  & $-0.017$ \tabularnewline
 & $g_{3}(0)$  & $-0.069$  & - -  &  & $g_{3}(0)$  & $-0.377$  & - -\tabularnewline
\hline 
$\Xi_{bc}^{+}\to\Lambda_{b}^{0}$  & $f_{1}(0)$  & $-0.45$  & $-0.55$  & $\Xi_{bc}^{+}\to\Sigma_{b}^{0}$  & $f_{1}(0)$  & $-0.23$  & $-0.32$ \tabularnewline
 & $f_{2}(0)$  & $0.31$  & $0.30$  &  & $f_{2}(0)$  & $1.51$  & $1.54$ \tabularnewline
 & $f_{3}(0)$  & $1.21$  & - -  &  & $f_{3}(0)$  & $0.38$  & - -\tabularnewline
 & $g_{1}(0)$  & $0.16$  & $-0.15$  &  & $g_{1}(0)$  & $0.33$  & $-0.41$\tabularnewline
 & $g_{2}(0)$  & $-0.14$  & $0.10$  &  & $g_{2}(0)$  & $-0.39$  & $0.18$ \tabularnewline
 & $g_{3}(0)$  & $-2.74$  & - -  &  & $g_{3}(0)$  & $-8.24$  & - -\tabularnewline
\hline 
$\Xi_{bc}^{0}\to\Lambda_{c}^{+}$  & $f_{1}(0)$  & $-0.141$  & $-0.113$  & $\Xi_{bc}^{0}\to\Sigma_{c}^{+}$  & $f_{1}(0)$  & $-0.069$  & $-0.071$ \tabularnewline
 & $f_{2}(0)$  & $-0.040$  & $-0.030$  &  & $f_{2}(0)$  & $0.159$  & $0.098$ \tabularnewline
 & $f_{3}(0)$  & $0.108$  & - -  &  & $f_{3}(0)$  & $-0.036$  & - -\tabularnewline
 & $g_{1}(0)$  & $0.067$  & $-0.047$  &  & $g_{1}(0)$  & $0.088$  & $-0.103$ \tabularnewline
 & $g_{2}(0)$  & $-0.037$  & $0.021$  &  & $g_{2}(0)$  & $0.059$  & $-0.003$ \tabularnewline
 & $g_{3}(0)$  & $-0.006$  & - -  &  & $g_{3}(0)$  & $-0.257$  & - -\tabularnewline
\hline 
\end{tabular}
\end{table}

A comparison between this work and other works in the literature can be found in Tables \ref{Tab:comparison_ff_cc} and \ref{Tab:comparison_ff_bb_bc} 
for the $cc$ sector, the $bb$ sector and the $bc$ sector with $c$
or $b$ quark decay.

Some comments: 
\begin{itemize}

\item The signs of the form factors of $c\to d$ processes ($\Xi_{cc}^{++}(ccu)\to\Lambda_{c}^{+}(dcu)$
and $\Xi_{bc}^{+}(cbu)\to\Lambda_{b}^{0}(dbu)$) in the LFQM have
been flipped so that those of vector-current form factors are the
same as ours. This stems from the asymmetry of $u$ and $d$ in the
wave-function of $\Lambda_{Q}=(1/\sqrt{2})(ud-du)Q$ with $Q=c/b$
in the final state.

\item It can be seen from Tables \ref{Tab:comparison_ff_cc} and \ref{Tab:comparison_ff_bb_bc} that, most of our results are comparable with others in other literature up to a sign difference for the axial-vector current form factors. However, this will not affect our predictions on physical observables, see Sec. \ref{sec:phenomenological}.

\item The sign conventions for $f_{2}$ and $g_{2}$ in Refs.~\cite{PerezMarcial:1989yh,Carson:1985pi} are different from ours in Eq.~(\ref{eq:parameterization}).
 
\end{itemize}

\subsection{Uncertainties}
\label{subsec:Uncertainties}

In this subsection, we will investigate the dependence of the form
factors on the inputs. $\Xi_{cc}^{++}\to\Sigma_{c}^{+}$ is taken as
an example. In Table \ref{Tab:error_estimate}, we have considered all the error sources including those from the heavy quark masses, Borel parameter $T_{1}^{2}$, thresholds $s_{1}^{0}$ and $s_{2}^{0}$, condensate parameters, pole residues and masses of initial and final baryons. 
One can see that the uncertainty mainly comes from that of the heavy quark mass $m_{c}$. That is, the results of the QCD sum rules are sensitive to the choice of the heavy quark mass. Similar situations are also encountered in studying other properties of heavy hadrons using QCD sum rules. In principle, this can be cured by calculating the contributions from the radiation corrections, which is undoubtedly a great challenge in the application of QCD sum rules. In this work, we will have to be content with the leading order results.
Also note that the dependence of the form factors on Borel parameter $T_{1}^{2}$ is weak.

When all uncertainties are considered, from Table \ref{Tab:error_estimate}, the error estimates of the form
factors at $q^{2}=0$ for $\Xi_{cc}^{++}\to\Sigma_{c}^{+}$ transition turn out to be 
\begin{align}
 & f_{1}(0)=-0.30\pm0.10,\quad f_{2}(0)=1.05\pm0.44,\quad f_{3}(0)=0.10\pm0.06,\nonumber \\
 & g_{1}(0)=0.46\pm0.18,\quad g_{2}(0)=-0.09\pm0.06,\quad g_{3}(0)=-2.96\pm1.30.
\label{eq:ff_err}
\end{align}

\begin{table}
\caption{The error estimates of the form factors for $\Xi_{cc}^{++}\to\Sigma_{c}^{+}$. }
\label{Tab:error_estimate}
\begin{tabular}{c|c|c|cccccccccc}
\hline 
\multicolumn{1}{c}{} &  & Central value & $m_{c}$ & $s_{1}^{0}$ & $s_{2}^{0}$ & $T_{1}^{2}$ & $\lambda_{i}$ & $\lambda_{f}$ & $M_{1}^{-}$ & $M_{2}^{-}$ & $\langle\bar{q}q\rangle$ & $\langle\bar{q}g_{s}\sigma Gq\rangle$\tabularnewline
\hline 
$f_{1}$ & $f_{1}(0)$ & $-0.30$ & $0.07$ & $0.01$ & $0.02$ & $0.02$ & $0.05$ & $0.03$ & $0.01$ & $0.00$ & $0.02$ & $0.00$\tabularnewline
 & $m_{{\rm pole}}$ & $1.76$ & $0.18$ & $0.27$ & $0.36$ & $0.03$ & $0.00$ & $0.00$ & $0.01$ & $0.01$ & $0.04$ & $0.02$\tabularnewline
 & $\delta$ & $-0.65$ & $0.67$ & $0.43$ & $1.03$ & $0.21$ & $0.00$ & $0.00$ & $0.00$ & $0.01$ & $0.04$ & $0.04$\tabularnewline
\hline 
$f_{2}$ & $f_{2}(0)$ & $1.05$ & $0.38$ & $0.02$ & $0.05$ & $0.04$ & $0.16$ & $0.12$ & $0.00$ & $0.00$ & $0.08$ & $0.02$\tabularnewline
 & $m_{{\rm pole}}$ & $1.57$ & $0.01$ & $0.02$ & $0.03$ & $0.02$ & $0.00$ & $0.00$ & $0.00$ & $0.00$ & $0.01$ & $0.01$\tabularnewline
 & $\delta$ & $0.23$ & $0.01$ & $0.00$ & $0.10$ & $0.04$ & $0.00$ & $0.00$ & $0.00$ & $0.00$ & $0.00$ & $0.01$\tabularnewline
\hline 
$f_{3}$ & $f_{3}(0)$ & $0.10$ & $0.00$ & $0.02$ & $0.04$ & $0.02$ & $0.02$ & $0.01$ & $0.02$ & $0.02$ & $0.00$ & $0.01$\tabularnewline
 & $m_{{\rm pole}}$ & $1.00$ & $0.08$ & $0.04$ & $0.01$ & $0.02$ & $0.00$ & $0.00$ & $0.05$ & $0.00$ & $0.00$ & $0.01$\tabularnewline
 & $\delta$ & $0.78$ & $0.43$ & $0.58$ & $0.34$ & $0.10$ & $0.00$ & $0.00$ & $0.39$ & $0.07$ & $0.05$ & $0.15$\tabularnewline
\hline 
$g_{1}$ & $g_{1}(0)$ & $0.46$ & $0.15$ & $0.01$ & $0.03$ & $0.00$ & $0.07$ & $0.05$ & $0.01$ & $0.01$ & $0.03$ & $0.00$\tabularnewline
 & $m_{{\rm pole}}$ & $2.29$ & $0.00$ & $0.18$ & $0.04$ & $0.06$ & $0.00$ & $0.00$ & $0.08$ & $0.04$ & $0.07$ & $0.05$\tabularnewline
 & $\delta$ & $0.41$ & $0.09$ & $0.59$ & $0.19$ & $0.02$ & $0.00$ & $0.00$ & $0.10$ & $0.05$ & $0.03$ & $0.02$\tabularnewline
\hline 
$g_{2}$ & $g_{2}(0)$ & $-0.09$ & $0.01$ & $0.03$ & $0.05$ & $0.01$ & $0.01$ & $0.01$ & $0.01$ & $0.00$ & $0.00$ & $0.01$\tabularnewline
 & $m_{{\rm pole}}$ & $1.20$ & $0.06$ & $0.13$ & $0.10$ & $0.00$ & $0.00$ & $0.00$ & $0.01$ & $0.00$ & $0.00$ & $0.02$\tabularnewline
 & $\delta$ & $1.59$ & $0.92$ & $2.98$ & $1.00$ & $0.32$ & $0.00$ & $0.00$ & $0.39$ & $0.14$ & $0.19$ & $0.35$\tabularnewline
\hline 
$g_{3}$ & $g_{3}(0)$ & $-2.96$ & $1.13$ & $0.04$ & $0.13$ & $0.07$ & $0.46$ & $0.34$ & $0.00$ & $0.01$ & $0.21$ & $0.13$\tabularnewline
 & $m_{{\rm pole}}$ & $1.34$ & $0.05$ & $0.03$ & $0.00$ & $0.02$ & $0.00$ & $0.00$ & $0.00$ & $0.00$ & $0.01$ & $0.00$\tabularnewline
 & $\delta$ & $0.16$ & $0.04$ & $0.03$ & $0.02$ & $0.02$ & $0.00$ & $0.00$ & $0.00$ & $0.00$ & $0.00$ & $0.00$\tabularnewline
\hline 
\end{tabular}
\end{table}

\section{Phenomenological applications}
\label{sec:phenomenological}

In this section, results for form factors will be applied to calculate the partial widths of semileptonic decays. 

\subsection{Semi-leptonic decays}

The effective Hamiltonian for the semi-leptonic process reads 
\begin{eqnarray}
{\cal H}_{{\rm eff}} & = & \frac{G_{F}}{\sqrt{2}}\bigg(V_{cs}^{*}[\bar{s}\gamma_{\mu}(1-\gamma_{5})c][\bar{\nu}\gamma^{\mu}(1-\gamma_{5})l]+V_{cd}^{*}[\bar{d}\gamma_{\mu}(1-\gamma_{5})c][\bar{\nu}\gamma^{\mu}(1-\gamma_{5})l]\bigg)\nonumber \\
&  & +\frac{G_{F}}{\sqrt{2}}V_{ub}[\bar{u}\gamma_{\mu}(1-\gamma_{5})b][\bar{l}\gamma^{\mu}(1-\gamma_{5})\nu],
\end{eqnarray}
where $G_{F}$ is Fermi constant and $V_{cs,cd,ub}$ are Cabibbo-Kobayashi-Maskawa
(CKM) matrix elements.

The helicity amplitudes will be used in the calculation and  for the vector current and the axial-vector
current, they are given as follows: 
\begin{eqnarray}
H_{\frac{1}{2},0}^{V} & = & -i\frac{\sqrt{Q_{-}}}{\sqrt{q^{2}}}\left((M_{1}+M_{2})f_{1}-\frac{q^{2}}{M_{1}}f_{2}\right),\;\;\;
H_{\frac{1}{2},0}^{A} =  -i\frac{\sqrt{Q_{+}}}{\sqrt{q^{2}}}\left((M_{1}-M_{2})g_{1}+\frac{q^{2}}{M}g_{2}\right),\nonumber \\
H_{\frac{1}{2},1}^{V} & = & i\sqrt{2Q_{-}}\left(-f_{1}+\frac{M_{1}+M_{2}}{M}f_{2}\right),\;\;\;
H_{\frac{1}{2},1}^{A}  =  i\sqrt{2Q_{+}}\left(-g_{1}-\frac{M_{1}-M_{2}}{M_{1}}g_{2}\right),\nonumber \\
H_{\frac{1}{2},t}^{V} & = & -i\frac{\sqrt{Q_{+}}}{\sqrt{q^{2}}}\left((M_{1}-M_{2})f_{1}+\frac{q^{2}}{M_{1}}f_{3}\right),\;\;\;
H_{\frac{1}{2},t}^{A} =  -i\frac{\sqrt{Q_{-}}}{\sqrt{q^{2}}}\left((M_{1}+M_{2})g_{1}-\frac{q^{2}}{M_{1}}g_{3}\right),
\end{eqnarray}
where $Q_{\pm}=(M_1\pm M_2)^{2}-q^{2}$ and $M_{1(2)}$ is the mass of the initial (final) baryon. The   amplitudes for negative helicity  are given by 
\begin{equation}
H_{-\lambda_{2},-\lambda_{W}}^{V}=H_{\lambda_{2},\lambda_{W}}^{V}\quad\text{and}\quad H_{-\lambda_{2},-\lambda_{W}}^{A}=-H_{\lambda_{2},\lambda_{W}}^{A},
\end{equation}
where $\lambda_{2}$ and $\lambda_{W}$ denote the polarizations of
the final baryon and the intermediate $W$ boson, respectively. Then
the helicity amplitudes for the $V-A$ current are obtained as 
\begin{equation}
H_{\lambda_{2},\lambda_{W}}=H_{\lambda_{2},\lambda_{W}}^{V}-H_{\lambda_{2},\lambda_{W}}^{A}.
\end{equation}

Decay widths for ${\cal B}_{1}\to{\cal B}_{2}l\nu$ with the longitudinally and transversely  polarized   $l\nu$ pair are evaluated  as
\begin{align}
\frac{d\Gamma_{L}}{dq^{2}} & =\frac{G_{F}^{2}|V_{{\rm CKM}}|^{2}q^{2}\ p\ (1-\hat{m}_{l}^{2})^{2}}{384\pi^{3}M_{1}^{2}}\left((2+\hat{m}_{l}^{2})(|H_{-\frac{1}{2},0}|^{2}+|H_{\frac{1}{2},0}|^{2})+3\hat{m}_{l}^{2}(|H_{-\frac{1}{2},t}|^{2}+|H_{\frac{1}{2},t}|^{2})\right),\label{eq:longi-1}\\
\frac{d\Gamma_{T}}{dq^{2}} & =\frac{G_{F}^{2}|V_{{\rm CKM}}|^{2}q^{2}\ p\ (1-\hat{m}_{l}^{2})^{2}(2+\hat{m}_{l}^{2})}{384\pi^{3}M_{1}^{2}}(|H_{\frac{1}{2},1}|^{2}+|H_{-\frac{1}{2},-1}|^{2}),\label{eq:trans-1}
\end{align}
where $\hat{m}_{l}\equiv m_{l}/\sqrt{q^{2}}$, $p=\sqrt{Q_{+}Q_{-}}/(2M_{1})$
is the magnitude of three-momentum of ${\cal B}_{2}$ in the rest frame of ${\cal B}_{1}$. Integrating out the squared momentum transfer $q^{2}$, we obtain the total decay width: 
\begin{equation}
\Gamma=\int_{m_l^2}^{(M_{1}-M_{2})^{2}}dq^{2}\frac{d\Gamma}{dq^{2}},
\end{equation}
where
\begin{equation}
\frac{d\Gamma}{dq^{2}}=\frac{d\Gamma_{L}}{dq^{2}}+\frac{d\Gamma_{T}}{dq^{2}}.
\end{equation}

The 
Fermi constant and CKM matrix elements are taken from Particle Data Group~\cite{Olive:2016xmw,Tanabashi:2018oca}: 
\begin{align}
& G_{F}=1.166\times10^{-5}{\rm GeV}^{-2},\nonumber \\
& |V_{cd}|=0.225,\quad|V_{cs}|=0.974,\quad |V_{ub}|=0.00357.\label{eq:GFCKM}
\end{align}
The lifetimes of the doubly heavy baryons are given in Table \ref{Tab:lifetime}.
The integrated partial decay widths, ratios of $\Gamma_{L}/\Gamma_{T}$
and the corresponding branching fractions are calculated and results
are given in Tables \ref{Tab:semi_lep_cc}, \ref{Tab:semi_lep_bb},
\ref{Tab:semi_lep_bc_c} and \ref{Tab:semi_lep_bc_b} respectively.
A comparison of our results with those in the literature is presented
in Table~\ref{Tab:comparison_semi_lep}.

\begin{table}
\caption{Results for the semi-leptonic decays: the $cc$ sector. The lifetimes
of the initial baryons, which are used to derive the branching fractions,
can be found in Table \ref{Tab:lifetime}. Here $l=e/\mu$. Here we have only considered the uncertainties from the heavy quark masses.}
 \label{Tab:semi_lep_cc}%
\begin{tabular}{l|c|c|c}
\hline 
Channel  & $\Gamma/(10^{-14}\ {\rm GeV})$  & ${\cal B}/10^{-3}$  & $\Gamma_{L}/\Gamma_{T}$ \tabularnewline
\hline 
$\Xi_{cc}^{++}\to\Lambda_{c}^{+}l^{+}\nu_{l}$  & $0.76\pm0.37$  & $2.97\pm1.42$  & $8.5\pm4.4$\tabularnewline
$\Xi_{cc}^{++}\to\Xi_{c}^{+}l^{+}\nu_{l}$  & $7.72\pm3.70$  & $30.00\pm14.40$  & $9.4\pm5.2$\tabularnewline
$\Xi_{cc}^{+}\to\Xi_{c}^{0}l^{+}\nu_{l}$  & $7.72\pm3.70$  & $5.16\pm2.47$  & $9.4\pm5.2$\tabularnewline
$\Omega_{cc}^{+}\to\Xi_{c}^{0}l^{+}\nu_{l}$  & $0.61\pm0.28$  & $1.90\pm0.87$  & $8.6\pm4.6$\tabularnewline
\hline
$\Xi_{cc}^{++}\to\Sigma_{c}^{+}l^{+}\nu_{l}$  & $0.49\pm0.29$  & $1.92\pm1.13$  & $1.1\pm0.2$\tabularnewline
$\Xi_{cc}^{++}\to\Xi_{c}^{\prime+}l^{+}\nu_{l}$  & $5.31\pm3.52$  & $20.70\pm13.70$  & $1.3\pm0.2$\tabularnewline
$\Xi_{cc}^{+}\to\Sigma_{c}^{0}l^{+}\nu_{l}$  & $0.99\pm0.58$  & $0.66\pm0.39$  & $1.1\pm0.2$\tabularnewline
$\Xi_{cc}^{+}\to\Xi_{c}^{\prime0}l^{+}\nu_{l}$  & $5.31\pm3.52$  & $3.55\pm2.36$  & $1.3\pm0.2$\tabularnewline
$\Omega_{cc}^{+}\to\Xi_{c}^{\prime0}l^{+}\nu_{l}$  & $0.56\pm0.35$  & $1.76\pm1.10$  & $1.0\pm0.2$\tabularnewline
$\Omega_{cc}^{+}\to\Omega_{c}^{0}l^{+}\nu_{l}$  & $12.50\pm8.02$  & $39.00\pm25.10$  & $1.2\pm0.2$\tabularnewline
\hline 
\end{tabular}
\end{table}
\begin{table}
\caption{Same as Table \ref{Tab:semi_lep_cc} but for the $bb$ sector.}
\label{Tab:semi_lep_bb}%
\begin{tabular}{l|c|c|c|l|c|c|c}
\hline 
Channel  & $\Gamma/\ (10^{-17}{\rm GeV})$  & ${\cal B}/10^{-5}$  & $\Gamma_{L}/\Gamma_{T}$  & Channel  & $\Gamma/\ (10^{-17}{\rm GeV})$  & ${\cal B}/10^{-5}$  & $\Gamma_{L}/\Gamma_{T}$ \tabularnewline
\hline 
$\Xi_{bb}^{-}\to\Lambda_{b}^{0}l^{-}\bar{\nu}_{l}$  & $2.19\pm1.62$  & $1.23\pm0.91$  & $7.3\pm5.3$  & $\Xi_{bb}^{-}\to\Lambda_{b}^{0}\tau^{-}\bar{\nu}_{\tau}$  & $1.03\pm0.79$  & $0.58\pm0.44$  & $7.4\pm5.8$ \tabularnewline
$\Omega_{bb}^{-}\to\Xi_{b}^{0}l^{-}\bar{\nu}_{l}$  & $5.34\pm3.97$  & $6.49\pm4.83$  & $5.5\pm4.0$  & $\Omega_{bb}^{-}\to\Xi_{b}^{0}\tau^{-}\bar{\nu}_{\tau}$  & $3.05\pm2.33$  & $3.71\pm2.84$  & $5.9\pm4.7$ \tabularnewline
$\Xi_{bb}^{0}\to\Sigma_{b}^{+}l^{-}\bar{\nu}_{l}$  & $11.70\pm10.20$  & $6.58\pm5.73$  & $0.8\pm0.3$  & $\Xi_{bb}^{0}\to\Sigma_{b}^{+}\tau^{-}\bar{\nu}_{\tau}$  & $6.42\pm5.50$  & $3.61\pm3.09$  & $1.0\pm0.3$ \tabularnewline
$\Xi_{bb}^{-}\to\Sigma_{b}^{0}l^{-}\bar{\nu}_{l}$  & $5.85\pm5.09$  & $3.29\pm2.87$  & $0.8\pm0.3$  & $\Xi_{bb}^{-}\to\Sigma_{b}^{0}\tau^{-}\bar{\nu}_{\tau}$  & $3.21\pm2.75$  & $1.81\pm1.55$  & $1.0\pm0.3$ \tabularnewline
$\Omega_{bb}^{-}\to\Xi_{b}^{\prime0}l^{-}\bar{\nu}_{l}$  & $7.72\pm6.24$  & $9.39\pm7.59$  & $0.8\pm0.3$  & $\Omega_{bb}^{-}\to\Xi_{b}^{\prime0}\tau^{-}\bar{\nu}_{\tau}$  & $4.20\pm3.31$  & $5.10\pm4.02$  & $1.0\pm0.3$ \tabularnewline
\hline 
\end{tabular}
\end{table}
\begin{table}
\caption{Same as Table \ref{Tab:semi_lep_cc} but for the charm decay of bottom-charm
baryons.}
\label{Tab:semi_lep_bc_c}%
\begin{tabular}{l|c|c|c}
\hline 
Channel  & $\Gamma/\ (10^{-14}{\rm GeV})$  & ${\cal B}/10^{-3}$  & $\Gamma_{L}/\Gamma_{T}$ \tabularnewline
\hline 
$\Xi_{bc}^{+}\to\Lambda_{b}^{0}l^{+}\nu_{l}$  & $0.82\pm0.39$  & $3.04\pm1.46$  & $11.0\pm6.2$\tabularnewline
$\Xi_{bc}^{+}\to\Xi_{b}^{0}l^{+}\nu_{l}$  & $4.37\pm2.00$  & $16.20\pm7.40$  & $8.8\pm5.2$\tabularnewline
$\Xi_{bc}^{0}\to\Xi_{b}^{-}l^{+}\nu_{l}$  & $4.37\pm2.00$  & $6.18\pm2.82$  & $8.8\pm5.2$\tabularnewline
$\Omega_{bc}^{0}\to\Xi_{b}^{-}l^{+}\nu_{l}$  & $0.30\pm0.13$  & $1.01\pm0.45$  & $8.5\pm4.8$\tabularnewline
\hline
$\Xi_{bc}^{+}\to\Sigma_{b}^{0}l^{+}\nu_{l}$  & $0.22\pm0.15$  & $0.82\pm0.57$  & $1.5\pm0.5$\tabularnewline
$\Xi_{bc}^{+}\to\Xi_{b}^{\prime0}l^{+}\nu_{l}$  & $2.52\pm1.75$  & $9.34\pm6.50$  & $1.7\pm0.4$\tabularnewline
$\Xi_{bc}^{0}\to\Sigma_{b}^{-}l^{+}\nu_{l}$  & $0.44\pm0.31$  & $0.62\pm0.44$  & $1.5\pm0.5$\tabularnewline
$\Xi_{bc}^{0}\to\Xi_{b}^{\prime-}l^{+}\nu_{l}$  & $2.52\pm1.75$  & $3.56\pm2.48$  & $1.7\pm0.4$\tabularnewline
$\Omega_{bc}^{0}\to\Xi_{b}^{\prime-}l^{+}\nu_{l}$  & $0.20\pm0.13$  & $0.65\pm0.42$  & $1.4\pm0.3$\tabularnewline
$\Omega_{bc}^{0}\to\Omega_{b}^{-}l^{+}\nu_{l}$  & $4.20\pm2.89$  & $14.10\pm9.66$  & $1.5\pm0.3$\tabularnewline
\hline 
\end{tabular}
\end{table}
\begin{table}
\caption{Same as Table \ref{Tab:semi_lep_cc} but for the bottom decay of bottom-charm
baryons.}
\label{Tab:semi_lep_bc_b}%
\begin{tabular}{l|c|c|c|l|c|c|c}
\hline 
Channel  & $\Gamma/\ (10^{-17}{\rm GeV})$  & ${\cal B}/10^{-5}$  & $\Gamma_{L}/\Gamma_{T}$  & Channel  & $\Gamma/\ (10^{-17}{\rm GeV})$  & ${\cal B}/10^{-5}$  & $\Gamma_{L}/\Gamma_{T}$ \tabularnewline
\hline 
$\Xi_{bc}^{0}\to\Lambda_{c}^{+}l^{-}\bar{\nu}_{l}$  & $8.23\pm4.78$  & $1.16\pm0.68$  & $4.8\pm1.7$  & $\Xi_{bc}^{0}\to\Lambda_{c}^{+}\tau^{-}\bar{\nu}_{\tau}$  & $6.53\pm4.03$  & $0.92\pm0.57$  & $6.1\pm2.5$ \tabularnewline
$\Omega_{bc}^{0}\to\Xi_{c}^{+}l^{-}\bar{\nu}_{l}$  & $6.99\pm3.81$  & $2.34\pm1.27$  & $5.6\pm2.8$  & $\Omega_{bc}^{0}\to\Xi_{c}^{+}\tau^{-}\bar{\nu}_{\tau}$  & $4.27\pm2.49$  & $1.43\pm0.83$  & $5.8\pm3.3$ \tabularnewline
$\Xi_{bc}^{+}\to\Sigma_{c}^{++}l^{-}\bar{\nu}_{l}$  & $17.50\pm7.78$  & $6.47\pm2.88$  & $0.5\pm0.1$  & $\Xi_{bc}^{+}\to\Sigma_{c}^{++}\tau^{-}\bar{\nu}_{\tau}$  & $10.50\pm4.53$  & $3.91\pm1.68$  & $0.6\pm0.1$ \tabularnewline
$\Xi_{bc}^{0}\to\Sigma_{c}^{+}l^{-}\bar{\nu}_{l}$  & $8.73\pm3.89$  & $1.23\pm0.55$  & $0.5\pm0.1$  & $\Xi_{bc}^{0}\to\Sigma_{c}^{+}\tau^{-}\bar{\nu}_{\tau}$  & $5.27\pm2.26$  & $0.74\pm0.32$  & $0.6\pm0.1$ \tabularnewline
$\Omega_{bc}^{0}\to\Xi_{c}^{\prime+}l^{-}\bar{\nu}_{l}$  & $9.79\pm4.31$  & $3.28\pm1.44$  & $0.6\pm0.1$  & $\Omega_{bc}^{0}\to\Xi_{c}^{\prime+}\tau^{-}\bar{\nu}_{\tau}$  & $5.79\pm2.48$  & $1.94\pm0.83$  & $0.6\pm0.1$ \tabularnewline
\hline 
\end{tabular}
\end{table}
\begin{table}
\caption{The decay widths (in units of GeV) for the semi-leptonic decays obtained in this work are compared with those from
the light-front quark model (LFQM)~\cite{Wang:2017mqp}, the heavy
quark spin symmetry (HQSS)~\cite{Albertus:2012nd}, the nonrelativistic
quark model (NRQM) and the MIT bag model (MBM)~\cite{PerezMarcial:1989yh}.}
\label{Tab:comparison_semi_lep} %
\begin{tabular}{c|c|c|c|c|c}
\hline 
Channel  & This work  & LFQM~\cite{Wang:2017mqp}  & HQSS~\cite{Albertus:2012nd}  & NRQM~\cite{PerezMarcial:1989yh}  & MBM~\cite{PerezMarcial:1989yh}\tabularnewline
\hline 
$\Xi_{cc}^{++}\to\Lambda_{c}^{+}l^{+}\nu_{l}$  & $(7.6\pm 3.7)\times10^{-15}$  & $1.05\times10^{-14}$  & $3.20\times10^{-15}$  & $1.97\times10^{-15}$  & $1.32\times10^{-15}$\tabularnewline
$\Xi_{cc}^{++}\to\Sigma_{c}^{+}l^{+}\nu_{l}$  & $(4.9\pm 2.9)\times10^{-15}$  & $9.60\times10^{-15}$  & $5.22\times10^{-15}$  & $6.58\times10^{-15}$  & $2.63\times10^{-15}$\tabularnewline
\hline 
$\Xi_{bb}^{-}\to\Lambda_{b}^{0}l^{-}\bar{\nu}_{l}$  & $(2.19\pm 1.62)\times10^{-17}$ & $1.58\times10^{-17}$  & - -  & - -  & - -\tabularnewline
$\Xi_{bb}^{-}\to\Sigma_{b}^{0}l^{-}\bar{\nu}_{l}$  & $(5.85\pm 5.09)\times10^{-17}$ & $3.33\times10^{-17}$  & - -  & - -  & - -\tabularnewline
\hline 
$\Xi_{bc}^{+}\to\Lambda_{b}^{0}l^{+}\nu_{l}$  & $(8.2\pm 3.9)\times10^{-15}$ & $6.85\times10^{-15}$  & - -  & - -  & - -\tabularnewline
$\Xi_{bc}^{+}\to\Sigma_{b}^{0}l^{+}\nu_{l}$  & $(2.2\pm 1.5)\times10^{-15}$  & $4.63\times10^{-15}$  & - -  & - -  & - -\tabularnewline
\hline 
$\Xi_{bc}^{0}\to\Lambda_{c}^{+}l^{-}\bar{\nu}_{l}$  & $(8.23\pm 4.78)\times10^{-17}$  & $1.84\times10^{-17}$  & - -  & - -  & - -\tabularnewline
$\Xi_{bc}^{0}\to\Sigma_{c}^{+}l^{-}\bar{\nu}_{l}$  & $(8.73\pm 3.89)\times10^{-17}$  & $4.74\times10^{-17}$  & - -  & - -  & - -\tabularnewline
\hline 
\end{tabular}
\end{table}

A few remarks are given in order. 
\begin{itemize}

\item The $c\to s$ induced channels like $\Xi_{cc}^{++}\to \Xi_{c}^+l^+\nu_l$ have a large branching fraction, typically at a few percent level.  This is comparable with the branching ratio of semileptonic $D$ decays~\cite{Olive:2016xmw,Tanabashi:2018oca}. 

\item Compared with Ref.~\cite{Wang:2017mqp}, in this work we have considered the contributions from the form factors $f_{3}$ and $g_{3}$.

\item  In the flavor SU(3) limit, there exist the following relations for the charm quark decay widths: 
\begin{align*}
\Gamma(\Xi_{cc}^{++}\to\Lambda_{c}^{+}l^{+}\nu) & =\Gamma(\Omega_{cc}^{+}\to\Xi_{c}^{0}l^{+}\nu),\;\;\;
\Gamma(\Xi_{cc}^{++}\to\Xi_{c}^{+}l^{+}\nu)  =\Gamma(\Xi_{cc}^{+}\to\Xi_{c}^{0}l^{+}\nu),\\
\Gamma(\Xi_{cc}^{++}\to\Sigma_{c}^{+}l^{+}\nu) & =\frac{1}{2}\Gamma(\Xi_{cc}^{+}\to\Sigma_{c}^{0}l^{+}\nu)=\Gamma(\Omega_{cc}^{+}\to\Xi_{c}^{\prime0}l^{+}\nu),\\
\Gamma(\Xi_{cc}^{++}\to\Xi_{c}^{\prime+}l^{+}\nu) & =\Gamma(\Xi_{cc}^{+}\to\Xi_{c}^{\prime0}l^{+}\nu)=\frac{1}{2}\Gamma(\Omega_{cc}^{+}\to\Omega_{c}^{0}l^{+}\nu),\\
\Gamma(\Xi_{bc}^{+}\to\Lambda_{b}^{0}l^{+}\nu) & =\Gamma(\Omega_{bc}^{0}\to\Xi_{b}^{-}l^{+}\nu),\;\;\;
\Gamma(\Xi_{bc}^{+}\to\Xi_{b}^{0}l^{+}\nu) =\Gamma(\Xi_{bc}^{0}\to\Xi_{b}^{-}l^{+}\nu),\\
\Gamma(\Xi_{bc}^{+}\to\Sigma_{b}^{0}l^{+}\nu) & =\frac{1}{2}\Gamma(\Xi_{bc}^{0}\to\Sigma_{b}^{-}l^{+}\nu)=\Gamma(\Omega_{bc}^{0}\to\Xi_{b}^{\prime-}l^{+}\nu),\\
\Gamma(\Xi_{bc}^{+}\to\Xi_{b}^{\prime0}l^{+}\nu) & =\Gamma(\Xi_{bc}^{0}\to\Xi_{b}^{\prime-}l^{+}\nu)=\frac{1}{2}\Gamma(\Omega_{bc}^{0}\to\Omega_{b}^{-}l^{+}\nu). 
\end{align*}
For the bottom quark decay,   the relations  for the decay widths are given as: 
\begin{align*} 
\Gamma(\Xi_{bb}^{-}\to\Lambda_{b}^{0}l^{-}\bar{\nu}) & =\Gamma(\Omega_{bb}^{-}\to\Xi_{b}^{0}l^{-}\bar{\nu}),\\
\Gamma(\Xi_{bb}^{0}\to\Sigma_{b}^{+}l^{-}\bar{\nu}) & =2\Gamma(\Xi_{bb}^{-}\to\Sigma_{b}^{0}l^{-}\bar{\nu})=2\Gamma(\Omega_{bb}^{-}\to\Xi_{b}^{\prime0}l^{-}\bar{\nu}), \\
\Gamma(\Xi_{bc}^{+}\to\Sigma_{c}^{++}l^{-}\bar{\nu}) & =2\Gamma(\Xi_{bc}^{0}\to\Sigma_{c}^{+}l^{-}\bar{\nu})=2\Gamma(\Omega_{bc}^{0}\to\Xi_{c}^{\prime+}l^{-}\bar{\nu}).
\end{align*}
Based on the results in Tables~\ref{Tab:semi_lep_cc}, \ref{Tab:semi_lep_bb},
\ref{Tab:semi_lep_bc_c}, and \ref{Tab:semi_lep_bc_b}, we find that
the SU(3) relations for some channels involving $\Omega_{bc}$ and $\Omega_{bb}$ are significantly broken. 

\item In Tables~\ref{Tab:semi_lep_cc}, \ref{Tab:semi_lep_bb}, \ref{Tab:semi_lep_bc_c}, and \ref{Tab:semi_lep_bc_b}, we have also shown the uncertainties for the phenomenological observables, which come from the uncertainties of $F(0)$'s of the corresponding form factors. The latter uncertainties in turn come from those of the heavy quark masses. In Subsetion \ref{subsec:Uncertainties}, we have seen that the uncertainty from the heavy quark mass dominates. 

\item It can be seen from Table \ref{Tab:comparison_semi_lep} that, most results in this work are comparable with those in the  literature.  
\end{itemize}

\subsection{Dependence of decay width on the form factors}

In this subsection, we will investigate the dependence of decay width
on the form factors taking $\Xi_{cc}^{++}\to\Sigma_{c}^{+}l^{+}\nu_{l}$ as an
example. The uncertainties of the decay width caused by those of the form factors in Eq. (\ref{eq:ff_err})
can be found in Table \ref{Tab:ff_width}. One can see that these uncertainties are quite different, of which the largest one comes from that of $g_{1}$. In fact, both $f_{3}$ and $g_{3}$ do not contribute to the decay width. This
is because the leptonic part of the amplitude $\bar{\nu}\gamma_{\mu}(1-\gamma_{5})l$
when contracted with $q^{\mu}$ from the hadronic matrix element vanishes if we neglect the masses of leptons. Finally, it is worth mentioning again that the uncertainty
of $g_{1}$ mainly comes from that of $m_{c}$, as can be seen from
Table \ref{Tab:error_estimate}.

The decay width turns out to be:
\begin{equation}
\Gamma(\Xi_{cc}^{++}\to\Sigma_{c}^{+}l^{+}\nu_{l}) =(4.94\pm3.51)\times10^{-15}\ {\rm GeV}.
\end{equation}
Here we have only considered the uncertainties from $F(0)$'s, and we have also checked that those
from $m_{{\rm pole}}$ and $\delta$ can be neglected. 
Note that here the uncertainties from $F(0)$'s include those from the heavy quark mass $m_{c}$, Borel parameter $T_{1}^{2}$, thresholds $s_{1}^{0}$ and $s_{2}^{0}$, condensate parameters, pole residues and masses of initial and final baryons. If we only consider the uncertainty from the heavy quark mass $m_{c}$ for $F(0)$'s, a slightly smaller error is obtained
\begin{equation}
\Gamma(\Xi_{cc}^{++}\to\Sigma_{c}^{+}l^{+}\nu_{l}) =(4.94\pm2.92)\times10^{-15}\ {\rm GeV}.
\end{equation}
It can be seen that, it is a good error estimate for the decay width if we only consider the uncertainties from the heavy quark masses. Thus, in Tables~\ref{Tab:semi_lep_cc}, \ref{Tab:semi_lep_bb}, \ref{Tab:semi_lep_bc_c}, and \ref{Tab:semi_lep_bc_b}, only the uncertainties from the heavy quark masses are considered.

\begin{table}
\caption{The uncertainties of the decay widths of $\Xi_{cc}^{++}\to\Sigma_{c}^{+}l^{+}\nu_{l}$ caused by those of the form factors in Eq. (\ref{eq:ff_err}). The central value of the decay width is $4.94\times 10^{-15}$ GeV. }
\label{Tab:ff_width} %
\begin{tabular}{cccccc}
\hline 
$f_{1}$  & $f_{2}$  & $f_{3}$  & $g_{1}$  & $g_{2}$  & $g_{3}$ \tabularnewline
\hline 
$9\%$  & $14\%$  & $0$  & $69\%$  & $4\%$  & $0$ \tabularnewline
\hline 
\end{tabular}
\end{table}

\section{Conclusions}

\label{sec:conclusions}

Since the observation of doubly
charmed baryon $\Xi_{cc}^{++}$ reported by LHCb,  many theoretical investigations have been triggered  on the hadron spectroscopy and   on the weak decays
of the doubly heavy baryons, most of which are based on phenomenological  models rooted in QCD.   
In this work, we have presented a first QCD sum rules analysis of  the form factors for the doubly heavy baryon decays into  singly heavy baryon. We have included the perturbative contribution and  condensation contributions up to dimension 5. We have also estimated  the partial  contributions
from the gluon-gluon  condensate, and found  that these contributions are negligible.
These form factors are then used to study the semi-leptonic decays.   Future experimental measurements can examine these predictions and test the validity to apply QCDSR to doubly-heavy baryons. 

With the advances of new LHCb measurements in future and the under-design experimental
facilities, it is anticipated that more theoretical works of analyzing
weak decays of doubly-heavy baryons will be conducted. In this direction,
we can foresee the following prospects. 
\begin{itemize}

\item In this study,  we have shown that part of the gluon-gluon condensate is small but an analysis with a complete estimate of gluon-gluon condensate  is  left for future. 

\item The interpolating currents for baryons are not uniquely determined.
An ideal option is to have a largest projection onto the ground state
of doubly-heavy baryons and to suppress the contributions from higher
resonances and continuum. The dependence on interpolating current
and an estimate of the corresponding uncertainties have to be conducted
in a systematic way.

\item Decay form factors calculated in this work are induced by   heavy to light transitions, and  the heavy to heavy transition will  be studied in future.  Another plausible framework is the non-relativistic QCD. 

\item We have investigated  the form factors defined by vector and axial-vector currents, while the  tensor form factor are necessary to study the flavor-changing neutral current  processes in bottom quark decays, like the radiative and the dilepton decay modes. 

\item We have focused on the final baryons  with  spin-1/2,  while the $1/2\to 3/2$ transition needs an independent analysis. 

\item  Our  calculation of the form factors is conducted at the leading order in the expansion of strong coupling constant. However, to achieve a more precise result, it is still necessary to perform the calculation of higher order radiative corrections in future works. 

\item The ordinary QCD sum rules makes use of small-$x$ OPE. In heavy to
light transition, there exists a large momentum transfer and it would
be advantageous to adopt the light-cone OPE. Recently, the authors of Ref. \cite{Shi:2019fph} conducted the light-cone QCDSR study, and similar results are obtained.

\end{itemize}

\section*{Acknowledgements}
The authors are grateful to Hai-Yang Cheng, Pietro Colangelo, J\"urgen K\"orner, Run-Hui
Li, Yu-Ming Wang, Zhi-Gang Wang, Fan-Rong Xu, Mao-Zhi Yang, Fu-Sheng
Yu for useful discussions. This work is supported in part by National
Natural Science Foundation of China under Grants No.11575110, 11735010,11911530088, 
Natural Science Foundation of Shanghai under Grants No.~15DZ2272100,
and by Key Laboratory for Particle Physics, Astrophysics and Cosmology,
Ministry of Education.


\begin{thebibliography}{10}  
 

\bibitem{Aaij:2017ueg} 
  R.~Aaij {\it et al.} [LHCb Collaboration],
  Phys.\ Rev.\ Lett.\  {\bf 119}, no. 11, 112001 (2017)
  doi:10.1103/PhysRevLett.119.112001
  [arXiv:1707.01621 [hep-ex]].


\bibitem{Aaij:2018wzf} 
  R.~Aaij {\it et al.} [LHCb Collaboration],
  Phys.\ Rev.\ Lett.\  {\bf 121}, no. 5, 052002 (2018)
  doi:10.1103/PhysRevLett.121.052002
  [arXiv:1806.02744 [hep-ex]].


\bibitem{Aaij:2018gfl} 
  R.~Aaij {\it et al.} [LHCb Collaboration],
  Phys.\ Rev.\ Lett.\  {\bf 121}, no. 16, 162002 (2018)
  doi:10.1103/PhysRevLett.121.162002
  [arXiv:1807.01919 [hep-ex]].


\bibitem{Traill:2017zbs} 
  M.~T.~Traill [LHCb Collaboration],
  PoS Hadron {\bf 2017}, 067 (2018).
  doi:10.22323/1.310.0067


\bibitem{Cerri:2018ypt} 
  A.~Cerri {\it et al.},
  arXiv:1812.07638 [hep-ph].


\bibitem{Wang:2017mqp} 
  W.~Wang, F.~S.~Yu and Z.~X.~Zhao,
  Eur.\ Phys.\ J.\ C {\bf 77}, no. 11, 781 (2017)
  doi:10.1140/epjc/s10052-017-5360-1
  [arXiv:1707.02834 [hep-ph]].


\bibitem{Meng:2017udf} 
  L.~Meng, N.~Li and S.~l.~Zhu,
  Eur.\ Phys.\ J.\ A {\bf 54}, no. 9, 143 (2018)
  doi:10.1140/epja/i2018-12578-2
  [arXiv:1707.03598 [hep-ph]].


\bibitem{Wang:2017azm} 
  W.~Wang, Z.~P.~Xing and J.~Xu,
  Eur.\ Phys.\ J.\ C {\bf 77}, no. 11, 800 (2017)
  doi:10.1140/epjc/s10052-017-5363-y
  [arXiv:1707.06570 [hep-ph]].


\bibitem{Gutsche:2017hux} 
  T.~Gutsche, M.~A.~Ivanov, J.~G.~Körner and V.~E.~Lyubovitskij,
  Phys.\ Rev.\ D {\bf 96}, no. 5, 054013 (2017)
  doi:10.1103/PhysRevD.96.054013
  [arXiv:1708.00703 [hep-ph]].


\bibitem{Li:2017pxa} 
  H.~S.~Li, L.~Meng, Z.~W.~Liu and S.~L.~Zhu,
  Phys.\ Lett.\ B {\bf 777}, 169 (2018)
  doi:10.1016/j.physletb.2017.12.031
  [arXiv:1708.03620 [hep-ph]].


\bibitem{Guo:2017vcf} 
  Z.~H.~Guo,
  Phys.\ Rev.\ D {\bf 96}, no. 7, 074004 (2017)
  doi:10.1103/PhysRevD.96.074004
  [arXiv:1708.04145 [hep-ph]].


\bibitem{Lu:2017meb} 
  Q.~F.~Lü, K.~L.~Wang, L.~Y.~Xiao and X.~H.~Zhong,
  Phys.\ Rev.\ D {\bf 96}, no. 11, 114006 (2017)
  doi:10.1103/PhysRevD.96.114006
  [arXiv:1708.04468 [hep-ph]].


\bibitem{Xiao:2017udy} 
  L.~Y.~Xiao, K.~L.~Wang, Q.~f.~Lu, X.~H.~Zhong and S.~L.~Zhu,
  Phys.\ Rev.\ D {\bf 96}, no. 9, 094005 (2017)
  doi:10.1103/PhysRevD.96.094005
  [arXiv:1708.04384 [hep-ph]].


\bibitem{Sharma:2017txj} 
  N.~Sharma and R.~Dhir,
  Phys.\ Rev.\ D {\bf 96}, no. 11, 113006 (2017)
  doi:10.1103/PhysRevD.96.113006
  [arXiv:1709.08217 [hep-ph]].


\bibitem{Ma:2017nik} 
  Y.~L.~Ma and M.~Harada,
  J.\ Phys.\ G {\bf 45}, no. 7, 075006 (2018)
  doi:10.1088/1361-6471/aac86e
  [arXiv:1709.09746 [hep-ph]].


\bibitem{Yu:2017zst} 
  F.~S.~Yu, H.~Y.~Jiang, R.~H.~Li, C.~D.~Lü, W.~Wang and Z.~X.~Zhao,
  Chin.\ Phys.\ C {\bf 42}, no. 5, 051001 (2018)
  doi:10.1088/1674-1137/42/5/051001
  [arXiv:1703.09086 [hep-ph]].


\bibitem{Meng:2017dni} 
  L.~Meng, H.~S.~Li, Z.~W.~Liu and S.~L.~Zhu,
  Eur.\ Phys.\ J.\ C {\bf 77}, no. 12, 869 (2017)
  doi:10.1140/epjc/s10052-017-5447-8
  [arXiv:1710.08283 [hep-ph]].


\bibitem{Hu:2017dzi} 
  X.~H.~Hu, Y.~L.~Shen, W.~Wang and Z.~X.~Zhao,
  Chin.\ Phys.\ C {\bf 42}, no. 12, 123102 (2018)
  doi:10.1088/1674-1137/42/12/123102
  [arXiv:1711.10289 [hep-ph]].


\bibitem{Cui:2017udv} 
  E.~L.~Cui, H.~X.~Chen, W.~Chen, X.~Liu and S.~L.~Zhu,
  Phys.\ Rev.\ D {\bf 97}, no. 3, 034018 (2018)
  doi:10.1103/PhysRevD.97.034018
  [arXiv:1712.03615 [hep-ph]].


\bibitem{Shi:2017dto} 
  Y.~J.~Shi, W.~Wang, Y.~Xing and J.~Xu,
  Eur.\ Phys.\ J.\ C {\bf 78}, no. 1, 56 (2018)
  doi:10.1140/epjc/s10052-018-5532-7
  [arXiv:1712.03830 [hep-ph]].


\bibitem{Xiao:2017dly} 
  L.~Y.~Xiao, Q.~F.~Lü and S.~L.~Zhu,
  Phys.\ Rev.\ D {\bf 97}, no. 7, 074005 (2018)
  doi:10.1103/PhysRevD.97.074005
  [arXiv:1712.07295 [hep-ph]].


\bibitem{Yao:2018zze} 
  X.~Yao and B.~Müller,
  Phys.\ Rev.\ D {\bf 97}, no. 7, 074003 (2018)
  doi:10.1103/PhysRevD.97.074003
  [arXiv:1801.02652 [hep-ph]].


\bibitem{Yao:2018ifh} 
  D.~L.~Yao,
  Phys.\ Rev.\ D {\bf 97}, no. 3, 034012 (2018)
  doi:10.1103/PhysRevD.97.034012
  [arXiv:1801.09462 [hep-ph]].


\bibitem{Ozdem:2018uue} 
  U.~Özdem,
  J.\ Phys.\ G {\bf 46}, no. 3, 035003 (2019)
  doi:10.1088/1361-6471/aafffc
  [arXiv:1804.10921 [hep-ph]].


\bibitem{Ali:2018ifm} 
  A.~Ali, A.~Y.~Parkhomenko, Q.~Qin and W.~Wang,
  Phys.\ Lett.\ B {\bf 782}, 412 (2018)
  doi:10.1016/j.physletb.2018.05.055
  [arXiv:1805.02535 [hep-ph]].


\bibitem{Dias:2018qhp} 
  J.~M.~Dias, V.~R.~Debastiani, J.-J.~Xie and E.~Oset,
  Phys.\ Rev.\ D {\bf 98}, no. 9, 094017 (2018)
  doi:10.1103/PhysRevD.98.094017
  [arXiv:1805.03286 [hep-ph]].


\bibitem{Li:2018epz} 
  R.~H.~Li and C.~D.~Lu,
  arXiv:1805.09064 [hep-ph].


\bibitem{Zhao:2018mrg} 
  Z.~X.~Zhao,
  Eur.\ Phys.\ J.\ C {\bf 78}, no. 9, 756 (2018)
  doi:10.1140/epjc/s10052-018-6213-2
  [arXiv:1805.10878 [hep-ph]].


\bibitem{Xing:2018bqt} 
  Y.~Xing and R.~Zhu,
  Phys.\ Rev.\ D {\bf 98}, no. 5, 053005 (2018)
  doi:10.1103/PhysRevD.98.053005
  [arXiv:1806.01659 [hep-ph]].


\bibitem{Zhu:2018epc} 
  R.~Zhu, X.~L.~Han, Y.~Ma and Z.~J.~Xiao,
  Eur.\ Phys.\ J.\ C {\bf 78}, 740 (2018)
  doi:10.1140/epjc/s10052-018-6214-1
  [arXiv:1806.06388 [hep-ph]].


\bibitem{Ali:2018xfq} 
  A.~Ali, Q.~Qin and W.~Wang,
  Phys.\ Lett.\ B {\bf 785}, 605 (2018)
  doi:10.1016/j.physletb.2018.09.018
  [arXiv:1806.09288 [hep-ph]].


\bibitem{Liu:2018euh} 
  M.~Z.~Liu, Y.~Xiao and L.~S.~Geng,
  Phys.\ Rev.\ D {\bf 98}, no. 1, 014040 (2018)
  doi:10.1103/PhysRevD.98.014040
  [arXiv:1807.00912 [hep-ph]].


\bibitem{Xing:2018lre} 
  Z.~P.~Xing and Z.~X.~Zhao,
  Phys.\ Rev.\ D {\bf 98}, no. 5, 056002 (2018)
  doi:10.1103/PhysRevD.98.056002
  [arXiv:1807.03101 [hep-ph]].


\bibitem{Bediaga:2018lhg} 
  R.~Aaij {\it et al.} [LHCb Collaboration],
  arXiv:1808.08865.


\bibitem{Wang:2018duy} 
  W.~Wang and R.~Zhu,
  arXiv:1808.10830 [hep-ph].


\bibitem{Dhir:2018twm} 
  R.~Dhir and N.~Sharma,
  Eur.\ Phys.\ J.\ C {\bf 78}, no. 9, 743 (2018).
  doi:10.1140/epjc/s10052-018-6220-3


\bibitem{Berezhnoy:2018bde} 
  A.~V.~Berezhnoy, A.~K.~Likhoded and A.~V.~Luchinsky,
  Phys.\ Rev.\ D {\bf 98}, no. 11, 113004 (2018)
  doi:10.1103/PhysRevD.98.113004
  [arXiv:1809.10058 [hep-ph]].


\bibitem{Jiang:2018oak} 
  L.~J.~Jiang, B.~He and R.~H.~Li,
  Eur.\ Phys.\ J.\ C {\bf 78}, no. 11, 961 (2018)
  doi:10.1140/epjc/s10052-018-6445-1
  [arXiv:1810.00541 [hep-ph]].


\bibitem{Zhang:2018llc} 
  Q.~A.~Zhang,
  Eur.\ Phys.\ J.\ C {\bf 78}, no. 12, 1024 (2018)
  doi:10.1140/epjc/s10052-018-6481-x
  [arXiv:1811.02199 [hep-ph]].


\bibitem{Li:2018bkh} 
  G.~Li, X.~F.~Wang and Y.~Xing,
  Eur.\ Phys.\ J.\ C {\bf 79}, no. 3, 210 (2019)
  doi:10.1140/epjc/s10052-019-6729-0
  [arXiv:1811.03849 [hep-ph]].


\bibitem{Meng:2018zbl} 
  L.~Meng and S.~L.~Zhu,
  Phys.\ Rev.\ D {\bf 100}, no. 1, 014006 (2019)
  doi:10.1103/PhysRevD.100.014006
  [arXiv:1811.07320 [hep-ph]].


\bibitem{Gutsche:2018msz} 
  T.~Gutsche, M.~A.~Ivanov, J.~G.~Körner, V.~E.~Lyubovitskij and Z.~Tyulemissov,
  Phys.\ Rev.\ D {\bf 99}, no. 5, 056013 (2019)
  doi:10.1103/PhysRevD.99.056013
  [arXiv:1812.09212 [hep-ph]].


\bibitem{Onishchenko:2000wf} 
  A.~I.~Onishchenko,
  hep-ph/0006271.


\bibitem{Onishchenko:2000yp} 
  A.~I.~Onishchenko,
  hep-ph/0006295.


\bibitem{Kiselev:2001fw} 
  V.~V.~Kiselev and A.~K.~Likhoded,
  Phys.\ Usp.\  {\bf 45}, 455 (2002)
  [Usp.\ Fiz.\ Nauk {\bf 172}, 497 (2002)]
  doi:10.1070/PU2002v045n05ABEH000958
  [hep-ph/0103169].


\bibitem{Zhang:2008rt} 
  J.~R.~Zhang and M.~Q.~Huang,
  Phys.\ Rev.\ D {\bf 78}, 094007 (2008)
  doi:10.1103/PhysRevD.78.094007
  [arXiv:0810.5396 [hep-ph]].


\bibitem{Wang:2010hs} 
  Z.~G.~Wang,
  Eur.\ Phys.\ J.\ A {\bf 45}, 267 (2010)
  doi:10.1140/epja/i2010-11004-3
  [arXiv:1001.4693 [hep-ph]].


\bibitem{Wang:2010vn} 
  Z.~G.~Wang,
  Eur.\ Phys.\ J.\ C {\bf 68}, 459 (2010)
  doi:10.1140/epjc/s10052-010-1357-8
  [arXiv:1002.2471 [hep-ph]].


\bibitem{Wang:2010it} 
  Z.~G.~Wang,
  Eur.\ Phys.\ J.\ A {\bf 47}, 81 (2011)
  doi:10.1140/epja/i2011-11081-8
  [arXiv:1003.2838 [hep-ph]].


\bibitem{Ioffe:2005ym} 
  B.~L.~Ioffe,
  Prog.\ Part.\ Nucl.\ Phys.\  {\bf 56}, 232 (2006)
  doi:10.1016/j.ppnp.2005.05.001
  [hep-ph/0502148].


\bibitem{Colangelo:2000dp} 
  P.~Colangelo and A.~Khodjamirian,
  In *Shifman, M. (ed.): At the frontier of particle physics, vol. 3* 1495-1576
  doi:10.1142/9789812810458\_0033
  [hep-ph/0010175].


\bibitem{Olive:2016xmw} 
  C.~Patrignani {\it et al.} [Particle Data Group],
  Chin.\ Phys.\ C {\bf 40}, no. 10, 100001 (2016).
  doi:10.1088/1674-1137/40/10/100001


\bibitem{Tanabashi:2018oca} 
  M.~Tanabashi {\it et al.} [Particle Data Group],
  Phys.\ Rev.\ D {\bf 98}, no. 3, 030001 (2018).
  doi:10.1103/PhysRevD.98.030001


\bibitem{Wang:2010fq} 
  Z.~G.~Wang,
  Eur.\ Phys.\ J.\ C {\bf 68}, 479 (2010)
  doi:10.1140/epjc/s10052-010-1365-8
  [arXiv:1001.1652 [hep-ph]].


\bibitem{Wang:2009cr} 
  Z.~G.~Wang,
  Phys.\ Lett.\ B {\bf 685}, 59 (2010)
  doi:10.1016/j.physletb.2010.01.039
  [arXiv:0912.1648 [hep-ph]].


\bibitem{Brown:2014ena} 
  Z.~S.~Brown, W.~Detmold, S.~Meinel and K.~Orginos,
  Phys.\ Rev.\ D {\bf 90}, no. 9, 094507 (2014)
  doi:10.1103/PhysRevD.90.094507
  [arXiv:1409.0497 [hep-lat]].


\bibitem{Roberts:2007ni} 
  W.~Roberts and M.~Pervin,
  Int.\ J.\ Mod.\ Phys.\ A {\bf 23}, 2817 (2008)
  doi:10.1142/S0217751X08041219
  [arXiv:0711.2492 [nucl-th]].


\bibitem{Karliner:2014gca} 
  M.~Karliner and J.~L.~Rosner,
  Phys.\ Rev.\ D {\bf 90}, no. 9, 094007 (2014)
  doi:10.1103/PhysRevD.90.094007
  [arXiv:1408.5877 [hep-ph]].


\bibitem{Cheng:2018mwu} 
  H.~Y.~Cheng and Y.~L.~Shi,
  Phys.\ Rev.\ D {\bf 98}, no. 11, 113005 (2018)
  doi:10.1103/PhysRevD.98.113005
  [arXiv:1809.08102 [hep-ph]].


\bibitem{Wang:2012kw} 
  Z.~G.~Wang,
  Eur.\ Phys.\ J.\ A {\bf 49}, 131 (2013)
  doi:10.1140/epja/i2013-13131-7
  [arXiv:1203.6252 [hep-ph]].


\bibitem{Ball:1991bs} 
  P.~Ball, V.~M.~Braun and H.~G.~Dosch,
  Phys.\ Rev.\ D {\bf 44}, 3567 (1991).
  doi:10.1103/PhysRevD.44.3567


\bibitem{PerezMarcial:1989yh} 
  R.~Perez-Marcial, R.~Huerta, A.~Garcia and M.~Avila-Aoki,
  Phys.\ Rev.\ D {\bf 40}, 2955 (1989)
  Erratum: [Phys.\ Rev.\ D {\bf 44}, 2203 (1991)].
  doi:10.1103/PhysRevD.44.2203, 10.1103/PhysRevD.40.2955


\bibitem{Carson:1985pi} 
  L.~J.~Carson, R.~J.~Oakes and C.~R.~Willcox,
  Phys.\ Rev.\ D {\bf 33}, 1356 (1986).
  doi:10.1103/PhysRevD.33.1356


\bibitem{Albertus:2012nd} 
  C.~Albertus, E.~Hernandez and J.~Nieves,
  PoS QNP {\bf 2012}, 073 (2012)
  doi:10.22323/1.157.0073
  [arXiv:1206.5612 [hep-ph]].


\bibitem{Shi:2019fph} 
  Y.~J.~Shi, Y.~Xing and Z.~X.~Zhao,
  Eur.\ Phys.\ J.\ C {\bf 79}, no. 6, 501 (2019)
  doi:10.1140/epjc/s10052-019-7014-y
  [arXiv:1903.03921 [hep-ph]].
 
\end{thebibliography}
\end{document}